\documentclass[reprint,amsmath,amssymb,aps,pre,superscriptaddress]{revtex4-1}
\usepackage{amsmath,amsthm,amsfonts,amssymb,bm}
\usepackage{graphicx}
\DeclareMathOperator{\sign}{sign}
\usepackage{color}
\usepackage{graphicx,amsmath,color,amsfonts}
\usepackage{hyperref} 
\hypersetup{citecolor=blue,colorlinks=true} 

\newcommand{\dP}{P'}
\newcommand{\dPs}{P'^{*}}

\newcommand{\gdot}{\dot{\gamma}}
\newcommand{\edot}{\dot{\epsilon}}

\newcommand{\be}{\begin{equation}}
\newcommand{\ee}{\end{equation}}
\newcommand{\beqna}{\begin{eqnarray}}
\newcommand{\eeqna}{\end{eqnarray}}

\newcommand{\taur}{\tau_R}
\newcommand{\taud}{\tau_d}

\newcommand{\vecv}[1]{\bm{{#1}}}
\newcommand{\tens}[1]{\bm{{#1}}}

\newcommand{\visc}{\tens{W}} 

\newcommand{\xhat}{\hat{\vecv{x}}}

\newcommand{\zhat}{\hat{\vecv{z}}}

\newcommand{\smf}[1] {\textcolor{blue}{\{\small{SMF: #1}\}}}
\newcommand{\hjb}[1] {\textcolor{green}{\{\small{HJB: #1}\}}}

\definecolor{indigo}{RGB}{75,0,130}
\definecolor{orange}{RGB}{255,165,0}
\definecolor{maroon}{RGB}{128,0,0}

\begin{document}

\author{H. J. Barlow}
 \affiliation{Department of Physics, Durham University, Science Laboratories,
  South Road, Durham DH1 3LE, UK}
\author{E. J. Hemingway}
 \affiliation{Department of Physics, Durham University, Science Laboratories,
  South Road, Durham DH1 3LE, UK}
  \author{A. Clarke}
  \affiliation{Schlumberger Cambridge Research, Madingley Road, Cambridge, UK, CB3 0EL, UK}
\author{S. M. Fielding}
 \affiliation{Department of Physics, Durham University, Science Laboratories,
  South Road, Durham DH1 3LE, UK}
  
\title{Linear instability of shear thinning pressure driven channel flow}

\begin{abstract}

We study theoretically pressure driven planar channel flow of shear thinning viscoelastic fluids. Combining linear stability analysis and full nonlinear simulation, we study the instability of an initially one-dimensional base state to the growth of two-dimensional perturbations with wavevector in the flow direction. We do so within three widely used constitutive models: the microscopically motivated Rolie-Poly model, and the phenomenological Johnson-Segalman and White-Metzner models. In each model, we find instability when the degree of shear thinning exceeds some level characterised by the logarithmic slope of the flow curve at its shallowest point, $n=d\log\Sigma/d\log\gdot|_{\rm min}$. Specifically, we find instability for $n<n^*$, with $n^*\approx 0.21, 0.11$ and $0.30$ in the Rolie-Poly, Johnson-Segalman and White-Metzner  models respectively. Within each model, we show that the critical pressure drop for the onset of instability obeys a criterion expressed in terms of this degree of shear thinning, $n$, together with the derivative of the first normal stress with respect to shear stress. Both shear thinning and rapid variations in first normal stress across the channel are therefore key ingredients driving the instability. In the Rolie-Poly and Johnson-Segalman models, the underlying mechanism appears to involve the destabilisation of a quasi-interface that exists in each half of the channel, across which the normal stress varies rapidly. (The flow is not however shear banded in any parameter regime that we consider.) In the White-Metzner model, no such quasi-interface exists, but the criterion for instability nonetheless appears to follow the same form as in the Rolie-Poly and Johnson-Segalman models. This presents an outstanding puzzle concerning any possibly generic nature of the instability mechanism. We finally make some brief comments on the Giesekus model, which is rather different in its predictions from the other three.

\end{abstract}

\maketitle

\section{Introduction}

Viscoelastic fluids commonly show flow instabilities in which an initially simple flow state gives way to a more complicated one~\cite{Larson1992}. Such instabilities arise even in the creeping flow regime of low Reynolds number, with the fluid inertia playing little or no role. They stem instead from the nonlinear way in which the viscoelastic dynamics interacts with an imposed flow. Instability typically sets in beyond a critical value of the imposed flow rate scaled by the viscoelastic relaxation time, as expressed by the dimensionless Weissenberg number. For flows with curved streamlines, such instabilities are widespread. They include the viscoelastic Taylor-Couette instability~\cite{larson1990purely}, instability of viscoelastic cone-plate flow~\cite{olagunju1995instabilities}, and  viscoelastic turbulence in plate-plate flow~\cite{schiamberg2006transitional}. A criterion for instability based on  the Weissenberg number and the curvature of the flow streamlines was put forward in Ref.~\cite{pakdel1996elastic}.

Considerable work has also been devoted to the question of whether creeping viscoelastic flows that have  {\em straight} streamlines (in the initially simple flow state) are also unstable. For fluids that have a constant viscosity as a function of flow rate (ie, fluids that don't shear thin, or thicken), but that are nonetheless still viscoelastic,  a one-dimensional (1D) base state has been shown to be linearly stable against the growth of 2D perturbations with wavevector in the flow direction, $x$, in both boundary driven planar Couette~\cite{wilson1999structure} and pressure driven Poiseuille flow~\cite{ho1977stability}, within the Oldroyd B model. However, a nonlinear analysis within the same model suggests a possible sub-critical (finite amplitude) instability at high enough Weissenberg number, given perturbations of large enough amplitude~\cite{morozov2005subcritical}. Experiments testing this prediction can be found in Refs.~\cite{qin2017characterizing,pan2013nonlinear}. 

The present paper concerns the stability properties of shear thinning fluids (ie, fluids that show a decrease in viscosity with increasing shear rate), in a commonly studied flow geometry with straight streamlines: that of pressure driven planar channel flow. An earlier stability analysis performed within the shear thinning White-Metzner model predicted an initially 1D base state to be linearly  unstable to the growth of 2D perturbations with wavevector in the flow direction,  for fluids that shear thin strongly enough, at large enough imposed pressure drops~\cite{wilson1999instabilityWM,Wilson2015,Castillo2017}.  Instability has also been predicted in the shear thinning Giesekus  and Phan-Thien Tanner models~\cite{grillet2002stability} and also, very recently, in a model of thixoelastoviscoplastic flow~\cite{Castillo2018}. Instability of shear thinning polymeric solutions in pressure driven flow has recently been confirmed experimentally~\cite{Picaut2017,Bodiguel2015,Poole2016,bonn2011large}. Instabilities have also long been known to arise during extrusion of polymeric fluids~\cite{denn2001extrusion}, although often with wall slip involved: a phenomenon that we ignore here.

Motivated by these works, the present study set out to address several outstanding questions concerning the stability properties of shear thinning fluids in pressure driven flows. What level of shear thinning is in general required for instability to arise?  Is such an instability generic across all fluids and constitutive models that show (at least) this level of shear thinning? Is the required level of shear thinning independent of all other constitutive properties of the fluid, or does it also depend on (for example) normal stresses? For any fluid showing at least the required minimal level of shear thinning, what is the critical pressure drop required for instability? Can we uncover a criterion for this critical pressure drop in terms of easily measurable rheological signatures, thereby providing a practical guide for when an experimentalist can expect instability?  Is the mechanism of instability the same in all constitutive models? And is instability predicted by constitutive models that are built on a microscopic understanding of the underlying polymeric relaxation processes, or only in the phenomenological models studied to date?  As will be seen in what follows, some of these questions will prove to have relatively straightforward answers; others remain unresolved by this study.

To address these questions, we have performed an extensive linear stability analysis together with full nonlinear numerical simulations within three widely used constitutive models: the
microscopically motivated Rolie-Poly model~\cite{Likhtman2003}, and the phenomenoligical Johnson-Segalman~\cite{Johnson1977},
and White-Metzner~\cite{white1963development} models. We shall also make some comments about the  Giesekus model~\cite{giesekus1982simple}, which is rather different in its predictions from the other three, in our concluding discussion Sec.~\ref{sec:conclusions}. As noted above, Giesekus~\cite{grillet2002stability} and White-Metzner~\cite{wilson1999instabilityWM} have already been shown to predict instability of pressure driven shear thinning channel flow. The purpose of studying them again here is to try to elucidate the degree to which the predictions of the different models can -- or cannot -- be understood within a common framework.

All of the models considered, except White-Metzner, are capable of strong enough shear thinning to give shear banding. We restrict ourselves here to parameter regimes in which thinning is less pronounced, with no banding.

We find an initially 1D base state, in which the flow varies only in the flow-gradient direction, $y$, to indeed be linearly unstable to the onset of 2D perturbations with wavevector in the flow direction, $x$, in all four models. (In Giesekus and White-Metzner this finding simply confirms the earlier predictions; we believe the result to be new in Rolie-Poly and Johnson-Segalman.) Within Rolie-Poly, Johnson-Segalman and White-Metzner, we further perform a comprehensive study of the phase behaviour of this instability as a function of the model parameters and the imposed pressure drop. We thereby calculate the minimal degree of shear thinning  needed for instability, within each model, expressed in terms of the logarithmic slope of the flow curve of shear stress as a function of shear rate $\Sigma(\gdot)$. We find instability below a critical value, $n^*$, of the logarithmic slope of the flow curve at its shallowest point, $n=d\log\Sigma/d\log\gdot|_{\rm min}$ with $n^*\approx 0.21$ in Rolie-Poly, $n^*\approx 0.11$ in Johnson-Segalman, and (as calculated previously in Ref.~\cite{wilson1999instabilityWM}) $n^*\approx 0.30$ in White-Metzner. In the Giesekus model, in contrast, $n^*$ depends strongly on the solvent viscosity, and is furthermore much lower than in the other models, varying (with solvent viscosity) in the range $0$ to $0.04$. The values in the Rolie-Poly and White-Metzner models are broadly consistent with instability having been observed for the shear thinning exponents of $n=0.21$ and $n=0.19$ in the experimental studies of Refs.~\cite{Bodiguel2015} and~\cite{Poole2016} respectively. Only the White-Metzner model appears consistent with the presence of instability for  the value of $n=0.29$ in the experiments of Ref.~\cite{Picaut2017}.

Within each of the Rolie-Poly, Johnson-Segalman and White-Metzner models, we show that the critical pressure drop for the onset of instability (expressed in units of the fluid modulus and the channel width) follows, to a good or reasonable level of approximation, a scaling function expressed in terms of (i) the logarithmic slope of the flow curve at its shallowest point (i.e., at the point of minimum slope), and (ii) the maximum derivative of the the first normal stress with respect to the shear stress.  In each of these three models, the scaling function is characterised by three dimensionless fitting parameters. Once these fitting parameters have been determined (for any model), a reasonable level of scaling collapse of the critical pressure drop at onset is achieved  for the data obtained across all values of the model parameters, for that model.

Our results thereby show that both strong shear thinning and large variations in normal stress across the channel tend to predispose a flow to instability.  The criterion just described is only a partial success, however. In particular, the values of the three fitting parameters are different from model to model. We are therefore unable to provide a criterion that is truly universal. Ideally, these fitting parameters would be recast in terms of any additional relevant dimensionless constitutive properties, leading to a re-expressed set of fitting parameters that have the same values across all models. We have been unable to achieve that in this work. It remains an open future goal,  which may however be unattainable.

In the Rolie-Poly and Johnson-Segalman models, we give evidence suggesting that the mechanism of instability involves the destabilisation of a quasi-interface at some location in each half of the channel, across which the first normal stress varies rapidly as a function of position, and on either side of which the fluid flows with a different local viscosity. In this way, the instability closely resembles that arising at the interface between layered fluids~\cite{wilson1997short}. Indeed, an extreme form of shear thinning arises when the underlying constitutive curve $\Sigma(\gdot)$ (for states of homogeneous shear flow) is non-monotonic. The flow state as predicted by a 1D calculation then comprises bands of differing shear rates and normal stresses. This 1D banded state is known to be unstable to the formation of two dimensional perturbations with wavevector in the flow direction, with the instability driven by a jump in first normal stress across the interface between the bands~\cite{fielding2005linear,fielding2006nonlinear,fielding2010shear}. The instability found here in the Johnson-Segalman and Rolie-Poly models therefore resembles that of an interface between shear bands, but now in a regime where the flow is no-longer truly banded.

In the White-Metzner model, no such quasi-interface exists, and the eigenfunction characterising the instability is spread diffusely across the channel. An unresolved question is why the same form of criterion for instability onset (albeit with different values of the three fitting parameters) appears to hold in the  White-Metzner model as in the Rolie-Poly and Johnson-Segalman models.

The paper is structured as follows. In Sec.~\ref{sec:models} we introduce the constitutive models that we shall study. In Sec.~\ref{sec:geometry} we define the flow geometry to be considered, then discuss units and parameter values in Sec.~\ref{sec:units}. In Sec.~\ref{sec:methods} we outline our calculation methods, which include linear stability and nonlinear simulations. We then present our results in Sec.~\ref{sec:results}, for the Rolie-Poly, Johnson-Segalman and White-Metzner models in turn. Finally we present our conclusions and discuss outstanding challenges in Sec.~\ref{sec:conclusions}.

\section{Models}
\label{sec:models}

We assume the total stress $\tens{\Sigma}(\vecv{r},t)$ in a fluid element at position $\vecv{r}$ at time $t$ to comprise a viscoelastic contribution $\tens{\sigma}(\vecv{r},t)$ from the entangled polymer chains, a Newtonian contribution of viscosity $\eta$, and an isotropic pressure, with $\tens{I}$ the unit tensor:
\be
\tens{\Sigma} = \tens{\sigma}+ 2 \eta \tens{D} - p\tens{I}.
\label{eqn:total_stress_tensor}
\ee
The Newtonian part, $2 \eta\tens{D}$, may arise from a true solvent, or model any fast polymeric relaxation modes that are not ascribed to the viscoelastic stress. Here $\tens{D} = \frac{1}{2}(\tens{K} + \tens{K}^T)$
is the symmetric part of the velocity gradient tensor, $K_{\alpha\beta} =
\partial_{\beta}v_{\alpha}$, where $\tens{v}(\tens{r},t)$ is the fluid
velocity field.  Below we shall also use the antisymmetric tensor $\tens{\Omega} = \frac{1}{2}(\tens{K} - \tens{K}^T)$. The isotropic pressure $p(\tens{r},t)$ acts to ensure that the flow remains incompressible:
\be
\label{eqn:incomp}
\vecv{\nabla}\cdot\vecv{v} = 0.
\ee
Throughout we shall work in the limit of creeping flow, in which the condition of force balance states that the total stress tensor must be divergence free:
\be
\vecv{\nabla}\cdot\,\tens{\Sigma} = 0.
\label{eqn:force_balance}
\ee
Accordingly, any instabilities that we report stem not from inertial effects, but are of purely viscoelastic origin. 

We write the viscoelastic stress in terms of a constant elastic modulus $G$ and a molecular conformation tensor $\visc(\tens{r},t)$ that describes the deformation of the polymer chains relative to an isotropic undeformed state,
with $\tens{\sigma}=G\visc$ in the Rolie-poly model
and $\tens{\sigma}=G\left[\visc-\tens{I}\right]$ in all the other models
that we shall consider. 
The dynamics of $\tens{W}(\vecv{r},t)$ in flow is then prescribed by a
viscoelastic constitutive equation~\cite{larson2013constitutive}. We
shall consider four shear thinning constitutive models that are widely
used across the rheology literature: the Rolie-Poly
model~\cite{Likhtman2003}, the Johnson-Segalman
model~\cite{Johnson1977}, the Giesekus
model~\cite{giesekus1982simple}, and the White-Metzner
model~\cite{white1963development}. (We map out the phase behaviour of
three of these in detail, and comment only briefly on the Giesekus
model.)

Three of these models admit such strong shear thinning that they predict shear banding in some regions of their parameter space. However, all the results presented below are for parameter regimes in which shear thinning is more moderate, with the constitutive curve of shear stress as a function of shear rate, $\Sigma_{xy}(\gdot)$, being monotonic, precluding shear banding. Therefore, in contrast to studies of instabilities between shear bands~\cite{fielding2005linear}, we do not include a stress diffusion term in our constitutive equations.

\subsection{Rolie-Poly model}
\label{sec:RP}

The flow behaviour of an entangled polymeric fluid can be modelled at a microscopic level by considering the molecular dynamics of the constituent polymer chains. The basic notion is that any test chain of interest has its motion laterally constrained by entanglements with other chains.  In a mean field approach, the constraining influence of these entanglements is represented by an effective tube to which that test chain is (initially) confined~\cite{doi1988theory}.  The GLAMM model~\cite{Graham2003} provides a stochastic equation for the dynamics of such a test chain in its tube. This model is however computationally prohibitive to work with in fluid dynamical simulations. We shall therefore work with a simpler approximation, in which GLAMM is projected onto a single mode description~\cite{Likhtman2003}. The resulting rolie-poly (RP) model prescribes the dynamics of the polymer conformation tensor as follows:
\begin{widetext}
\beqna
\partial_t{\visc}+\tens{v}\cdot\nabla\tens{\visc} = \tens{K} \cdot \visc + \visc \cdot \tens{K}^T - \frac{1}{\taud}\left(\visc - \tens{I}\right) - \frac{2(1-A)}{\taur}\left[\, \visc + \beta A^{-2\delta}\left(\visc - \tens{I}\right) \right].
\label{eqn:rolie-poly_tensor_stretch}
\eeqna
\end{widetext}
In this equation, $\taud$ is the characteristic 
timescale on which a test chain escapes its tube of constraints via a process of 1D curvilinear diffusion along its own length, known as reptation. The Rouse
time $\taur$ is the much shorter timescale on which the degree of chain stretch
relaxes, by the lowest mode of curvilinear breathing of a chain within its tube. 
The chain stretch $A = \sqrt{3/T\,}$, where $T = \text{tr}\,\tens{\visc}$. The ratio $\taud/\taur$ of these two relaxation times increases with the number of entanglements per chain $Z$. We work throughout with highly entangled polymers and use the simpler, non-stretching version of the Rolie-Poly model, which follows from Eqn.~\ref{eqn:rolie-poly_tensor_stretch} in the limit $\taur\to 0$:
\begin{widetext}
\beqna
\partial_t{\visc}+\tens{v}\cdot\nabla\tens{\visc} = \tens{K} \cdot \visc + \visc \cdot \tens{K}^T - \frac{1}{\tau}\left(\visc - \tens{I}\right) - \dfrac{2}{3}\rm{Tr}(\tens{K}\cdot\tens{W})\left[\tens{W}+\beta(\tens{W}-\tens{I})\right].
\label{eqn:rolie-poly_tensor}
\eeqna
\end{widetext}
We thereby assume that flow rates, and the rates associated with the onset of any instability, remain modest compared with $1/\taur$.
Note that we now denote $\taud$ simply by $\tau$. The parameter $\beta$ describes the level of  ``convective constraint release" (CCR)~\cite{ianniruberto2014convective}, in
which the relaxation of polymer chain stretch also relaxes entanglement
points, thereby also allowing relaxation of tube orientation. It has a
range $0 \leq \beta \leq 1$, with no current consensus on its value within that range.  For $\beta<1$ the model is capable of capturing shear banding, at low values of the solvent viscosity $\eta$. In what follows, we consider only values of $\beta,\eta$ that are outside the banding regime.

\subsection{Johnson-Segalman model}

Alongside the Rolie-Roly model, which was derived by considering the
molecular relaxation processes of reptation, chain stretch relaxation
and CCR as just discussed, we shall also consider some constitutive
models that are instead phenomenological in origin.  We start with the
Johnson Segalman model~\cite{Johnson1977}. This is closely related to
the Oldroyd B model~\cite{spiess1987rb}, which considers the dynamics
of an ensemble of dumbbells in a solvent. Each dumbbell comprises two
beads connected by a spring, and is taken to model a single polymer
molecule. Each bead is subject to Brownian motion, Stokes drag, and
the spring force.  In the Oldroyd B model, each dumbbell is assumed to
deform in flow in an affine way, following the background solvent.  In
contrast, in the Johnson-Segalman model each dumbbell is instead taken
to slip relative to the solvent in a non-affine way. This phenomenon
is described by a slip parameter $a$, which obeys $|a|\le 1$. The
molecular conformation tensor then obeys:
\begin{widetext}
\beqna
\partial_t{\visc}+\tens{v}\cdot\nabla\tens{\visc} = (\tens{\Omega}\cdot\visc  - \visc\cdot\tens{\Omega}) + a(\tens{D} \cdot \visc + \visc \cdot \tens{D}) - \frac{1}{\tau}\left(\visc - \tens{I}\right),
\eeqna
\end{widetext}
with a relaxation time $\tau$.   For $|a|<1$, the model captures shear thinning. (For $\eta<0.125$, it further captures shear banding. We set $\eta>0.125$ in all of what follows.) For $a=1$, the Oldroyd B model is recovered, with no shear thinning. We note that the Johnson-Segalman model predicts unphysical time-dependent behaviour in shear startup at high strain rates. This pathology appears not to cause problems with the stability calculation here, however: even at high strain rates, the base state about which we linearise is stationary.

\subsection{White-Metzner model}
\label{sec:WM}

The White-Metzner model captures shear thinning by invoking a  relaxation time that depends on the frame-invariant strain rate, $\gdot\equiv\sqrt{\tens{D}:\tens{D}}$, as follows:
%
\beqna
\partial_t{\visc}+\tens{v}\cdot\nabla\tens{\visc} = \tens{K}\cdot \visc + \visc \cdot \tens{K}^T - \frac{1}{\tau^n\gdot^{n-1}}\left(\visc - \tens{I}\right).\nonumber
\label{eqn: WM_constit}
\eeqna
%
The power law index $0<n\le 1$, giving shear thinning for $n<1$, and recovering the  Oldroyd B model for $n=1$.

Note that the White-Mezner model gives power law fluid behaviour even in the limit of $\gdot\tau\to 0$, lacking any terminal Newtonian regime of weak shear. As such, it should be treated with some caution in modelling the experimental polymeric solutions that inspired this work, at least in the limit of low strain rates (which are always obtained near the centre of the channel, even at high pressure drops).

\subsection{Giesekus model}

Another phenomenological model commonly used to describe
concentrated polymeric solutions or melts is due to Giesekus~\cite{giesekus1982simple}. It considers an anisotropic drag on polymer chains that are oriented due to flow, and models this via an anisotropy parameter $\alpha$, which must lie in the range $0\leq \alpha\leq 1$. The molecular conformation evolves according to

\begin{widetext}
\beqna
\partial_t{\visc}+\tens{v}\cdot\nabla\tens{\visc} = \tens{K}\cdot \visc + \visc \cdot \tens{K}^T - \frac{1}{\tau}\left(\visc - \tens{I}\right) - \frac{\alpha}{\tau}\left(\visc - \tens{I}\right)^2,
\label{eqn: giesekus_constit}
\eeqna
\end{widetext}
with relaxation time $\tau$. For $\alpha>0.5$ the model is capable of capturing shear banding, at low enough values of $\eta$. We consider only $\alpha,\eta$ values that are outwith the banding regime in what follows. For $\alpha=0$, the Oldroyd B model is recovered.

\section{Flow Geometry}
\label{sec:geometry}

We consider a slab of fluid confined between flat parallel plates at $y=-L_y/2$ and $y=+L_y/2$. The flow is driven along the channel in the flow direction of positive $\xhat$ by means of a negative pressure gradient per unit length along $x$, of magnitude $\dP$. The flow is assumed to remain translationally invariant in the vorticity direction $\zhat$, and with no component of velocity in $\zhat$. (These assumptions are in accordance with Squire's theorem, but should be checked in future in fully 3D studies, in particular with respect to the potentially destabilising influence of second normal stress differences~\cite{LOCKETT1969337}. )   Boundary conditions of no-slip and no-permeation are assumed at the plates at $y=\pm L_y/2$. In the flow direction the channel has length $L_x$, with periodic boundary conditions. 

Note that we have chosen to use $\dP$ as the parameter controlling the strength of flow in this study. An alternative choice, commonly used in other studies~\cite{grillet2002stability,wilson1999instabilityWM}, would be to instead characterise the strength of flow by a Weissenberg number, defined by the shear rate at some point across the channel (or the velocity on the centreline divided by the channel width), multiplied by the fluid's viscoelastic relaxation time.

\section{Units and parameter values}
\label{sec:units}

We report all results below in units of length in which the channel width $L_y=1$; units of time in which the basic polymeric relaxation timescale $\tau=1$; and units of mass in which the polymeric shear modulus $G=1$.

In these units, there remain just four parameters to explore. For each of the constitutive models that we consider, the first is the single parameter pertaining to the viscoelastic dynamics: the convective constraint release parameter $\beta$ in the Rolie-Poly model, the slip parameter $a$ in the Johnson-Segalman model, 
the power law index $n$ in the White-Metzner model, and the anisotropy parameter $\alpha$ in the Giesekus model.
In any place where we need a single symbol to denote this viscoelastic parameter in a model-independent way, we use $\xi$. The second parameter is the Newtonian solvent viscosity $\eta$, which we take to be always small compared to the scale of the polymer viscosity $G\tau=1$. The third is the channel length $L_x$. In our linear stability analysis we report dispersion relations as a continuous function of the magnitude of the wavevector $q\xhat$. However, it is important to realise that this is actually quantized as $q=n\pi/L_x$ for any channel of finite length $L_x$. In our nonlinear simulations, $L_x$ is an explicit parameter. The final parameter is the pressure drop, $\dP$, characterising the strength of the imposed flow.

\section{Calculation Methods}
\label{sec:methods}

The flow models introduced in Sec.~\ref{sec:models} above all have the same basic structure, which we recap as follows. The total stress
\be
\tens{\Sigma} = \tens{\sigma}+ 2 \eta \tens{D} - p\tens{I},
\label{eqn:form1}
\ee
In creeping flow,  force balance requires:
\be
\vecv{\nabla}\cdot\,\tens{\Sigma} = 0.
\label{eqn:form2}
\ee
The condition of incompressibility gives:
\be
\label{eqn:form3}
\vecv{\nabla}\cdot\vecv{v} = 0.
\ee
The viscoelastic stress
\be
\label{eqn:form4}
\tens{\sigma}=\tens{\sigma}(\tens{W}), 
\ee
in which the molecular conformation tensor $\tens{W}$ evolves according to a viscoelastic constitutive model. Above, we specified several different constitutive models. However all have the same general form:
\be
\label{eqn:form5}
\partial_t{\visc}+\tens{v}\cdot\nabla\tens{\visc} = F(\visc,\nabla\vecv{v},\xi).
\ee
The parameter $\xi=\beta$ in the Rolie-Poly model, $\xi=a$ in the Johnson-Segalman model,  $\xi=n$ in White-Metzner, and $\xi=\alpha$ in Giesekus.
These equations must be solved subject to a constant imposed pressure gradient $-\dP\xhat$ per unit length in $x$.

\begin{figure*}[t!]
\vspace{-0.5cm}
\includegraphics[width=0.475\textwidth]{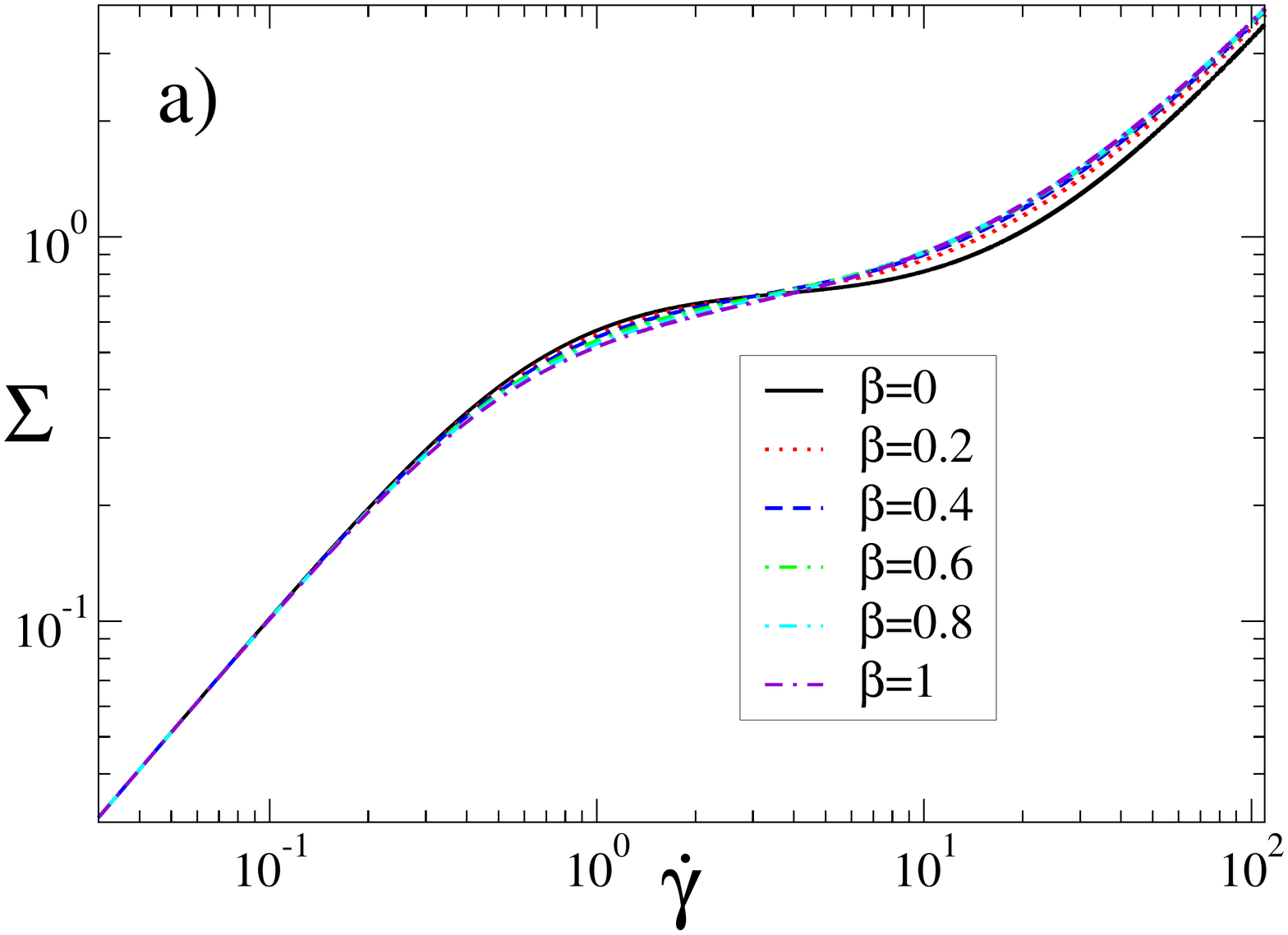}
\includegraphics[width=0.475\textwidth]{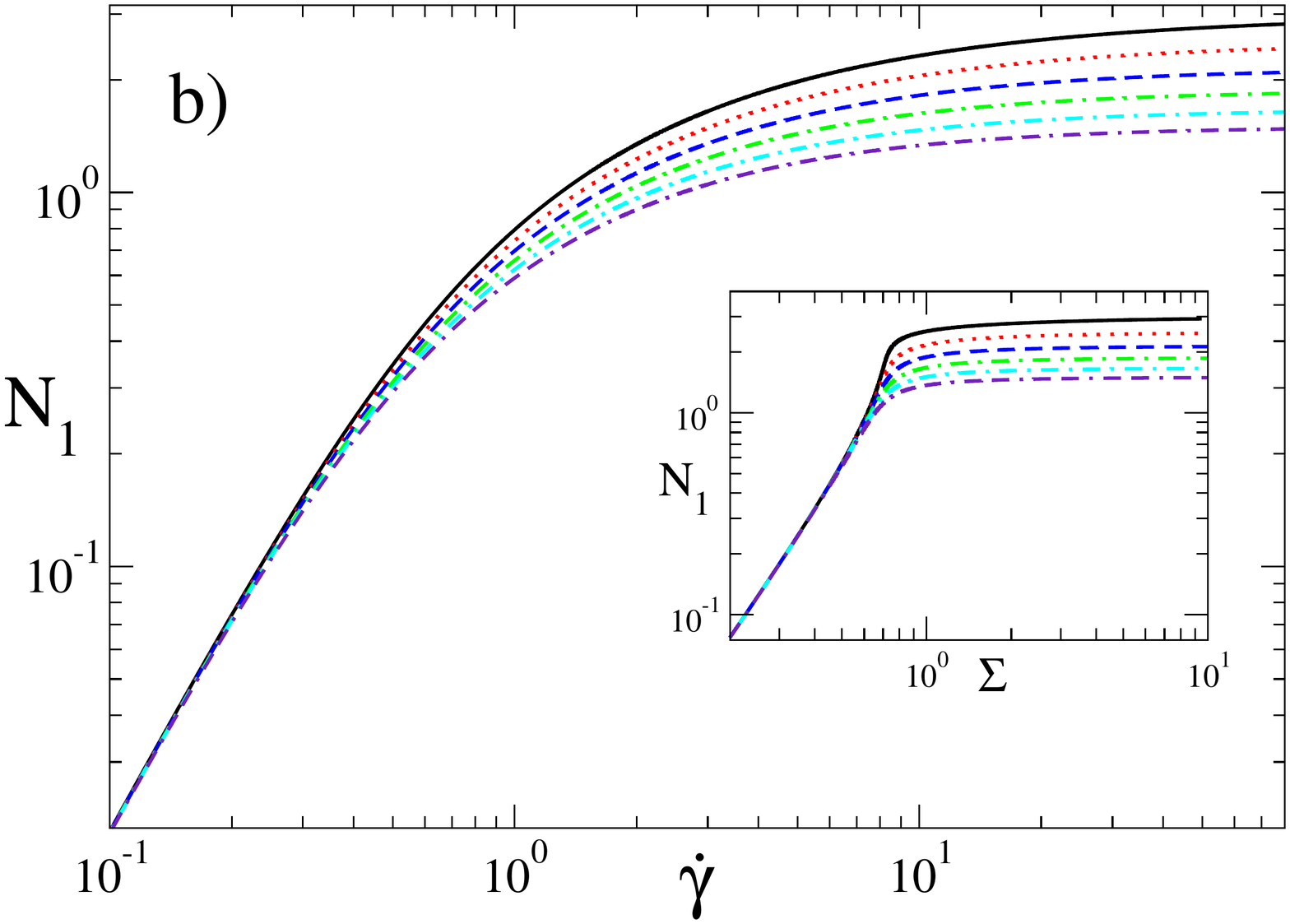}
\caption{(a) Flow curves of shear stress as a function of shear rate in homogeneous (0D) shear flow, computed in the Rolie-Poly model for  several values of the CCR parameter $\beta$, and a solvent viscosity $\eta=0.03$. (b) Corresponding normal stress as a function of strain rate and (inset) plotted parametrically as a function of shear stress.  }
\label{fig:RP_curves}
\end{figure*}

For any values of the model parameters $\xi,\eta$ and imposed pressure
gradient $\dP$, we first calculate the one-dimensional (1D) stationary
solution to Eqns.~\ref{eqn:form1} -~\ref{eqn:form5} that depends only
on the flow gradient direction $y$. (We do so by numerically
evolving the model equations to steady state, allowing spatial
variations only in $y$.) We denote this initial base state
$\vecv{v}_0(y)$, $\tens{\visc}_0(y)$, $\tens{\sigma}_0(y)$,
$\tens{\Sigma}_0(y)$, $p_0(y)$. (It also depends on $\xi,\eta$ and
$\dP$, though it would be cumbersome to write these dependencies
explicitly.)  The basic question that we now address is whether, for
any $\xi,\eta,\dP$ and channel length $L_x$, this 1D base state is
stable or unstable with respect to the growth of 2D perturbations that
depend on the flow direction $x$ as well as $y$.

We address this question first by performing a linear stability analysis, in which we add to the base state $\vecv{v}_0(y)$, $\tens{\visc}_0(y)$, $\tens{\sigma}_0(y)$, $\tens{\Sigma}_0(y)$, $p_0(y)$ small amplitude perturbations as follows:
\beqna
\vecv{v}(x,y,t)&=&\vecv{v}_0(y)+\sum_q\delta\vecv{v}_q(y)\exp(iqx+\omega_q t),\nonumber\\
\tens{\visc}(x,y,t)&=&\tens{\visc}_0(y)+\sum_q\delta\tens{\visc}_q(y)\exp(iqx+\omega_q t),\nonumber\\
\tens{\sigma}(x,y,t)&=&\tens{\sigma}_0(y)+\sum_q\delta\tens{\sigma}_q(y)\exp(iqx+\omega_q t),\nonumber\\
\tens{\Sigma}(x,y,t)&=&\tens{\Sigma}_0(y)+\sum_q\delta\tens{\Sigma}_q(y)\exp(iqx+\omega_q t),\nonumber\\
p(x,y,t)&=&p_0(y)+\sum_q\delta p_q(y)\exp(iqx+\omega_q t).\nonumber\\
\eeqna
Substituting these into the governing Eqns.~\ref{eqn:form1}
-~\ref{eqn:form5}, expanding in powers of the amplitude of the
perturbations, and retaining only terms of first order in that
amplitude, gives a set of linearised equations governing the dynamics
of the perturbations. These equations are valid in the linear regime
in which the perturbations remain small. In this linear regime, the
$q-$modes are all independent of each other. The main quantities of
interest are then, for each value of $q$, the eigenvalue $\omega_q$
with the largest real part, and the associated eigenfunction
$\delta\vecv{v}_q(y),\delta\tens{\visc}_q(y),\delta\tens{\sigma}_q(y),\delta\tens{\Sigma}_q(y),\delta
p_q(y)$. The sign of $\Re\omega_q$ then determines whether the
associated perturbation, with a functional form prescribed by the
eigenfunction, grows or decays. A positive value $\Re\omega_q>0$
signifies linear instability to the growth of a two-dimensional
perturbation $\propto \exp(iqx)$. In contrast, if $\Re\omega_q <0$ for
all modes, the base state is linearly stable.
\\The `fully linearised' recipe of the previous paragraph gives an exact
but cumbersome method for calculating the eigenmodes of the
instability, as just described. In calculational practice, however, we
used a different method, which is equivalent to that of the
previous paragraph in the early time linear regime, as long as the
perturbations remain small. Specifically, we write a `partly
linearised' viscoelastic constitutive equation, at any wavevector $q$,
as follows:
\\
\begin{widetext}
\begin{equation}
\label{eqn:formlin1}
\partial_t{\delta\visc}_q+i q u_0 \delta\,\tens{\visc}_q+\delta v_q\,\partial_y\tens{\visc}_0
=F(\visc_0+\delta\tens{\visc}_q,\nabla\vecv{v}_0+e^{-iqx}\nabla(\delta\vecv{v}_q e^{iqx}),\xi),
\end{equation}
\end{widetext}
and the force balance equation as
\begin{equation}
\eta\nabla^2\delta \tens{v}_q+\nabla\cdot\delta\tens{\Sigma}_q-\nabla\delta p_q=0.
\end{equation}
In Eqn.~\ref{eqn:formlin1}, $u_0$ is the base state speed, defined by $\tens{v}_0=u_0\hat{\tens{x}}$. $\delta v_q$ corresponds to the magnitude of the y-components of the perturbation velocity $\delta\tens{v}_q=\delta u_q\hat{\tens{x}}+\delta v_q \hat{\tens{y}}$.
Note that in Eqn.~\ref{eqn:formlin1}, we have linearised advective terms of the viscoelastic constitutive
equation exactly, but left the rest of the constitutive equation written in
nonlinear form.
Into these partly linearised equations we substitute the
time-independent base state $\tens{W}_0(y)$ and $\vecv{v}_0(y)$ as
calculated above, and initialise spatially random (in y) perturbations
$\delta\tens{W}_q$, $\delta\vecv{v}_q$, etc, with a small amplitude. We then time-step these equations and observe the
growth (or decay) of the perturbations in time. The slope of the logarithm of
the amplitude of the perturbations as a function of time then gives
the real part of the eigenvalue $\Re\omega_q$. (The imaginary part of
the eigenvalue instead merely represents the advection rate of the
perturbations~\cite{Castillo2018} and is not reported here.)  The flow
pattern that emerges then gives the associated eigenfunction. We have
checked this method against earlier calculations that used the fully
linearised method, for the case of shear banded flow~\cite{fielding2010shear}.

An important aim will be to relate these stability properties to the
functional form of the 1D base state profiles of shear rate,
$\gdot(y)$, shear stress, $\Sigma(y)$, and first normal stress,
$N_1(y)$; and via these to the functional form of the 0D flow curves
$\Sigma(\gdot)$ and $N_1(\gdot)$, computed (or measured) for a state
of homogeneous shear flow. In this way, we seek to give a practical
guide to when an experimentalist might expect a given pressure driven
flow to be stable or unstable. Note that we denote
$\Sigma_{xy}=\Sigma$ and drop the 0 subscript from the base state for
ease of notation. We adopt the usual definition of the first normal
stress, $N_1=\Sigma_{xx}-\Sigma_{yy}$.

\begin{figure*}[t!]
\includegraphics[width=0.475\textwidth]{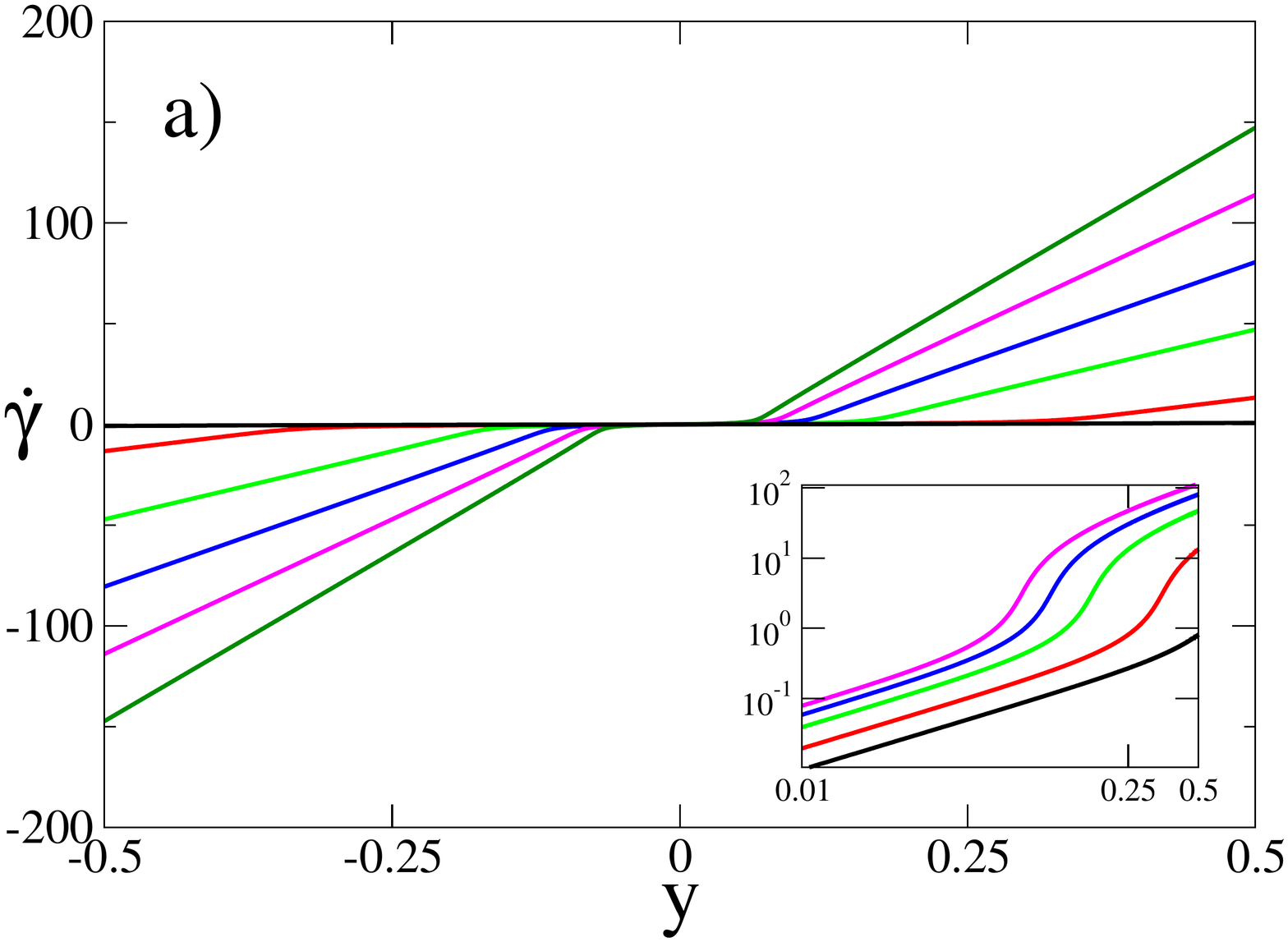}
\includegraphics[width=0.475\textwidth]{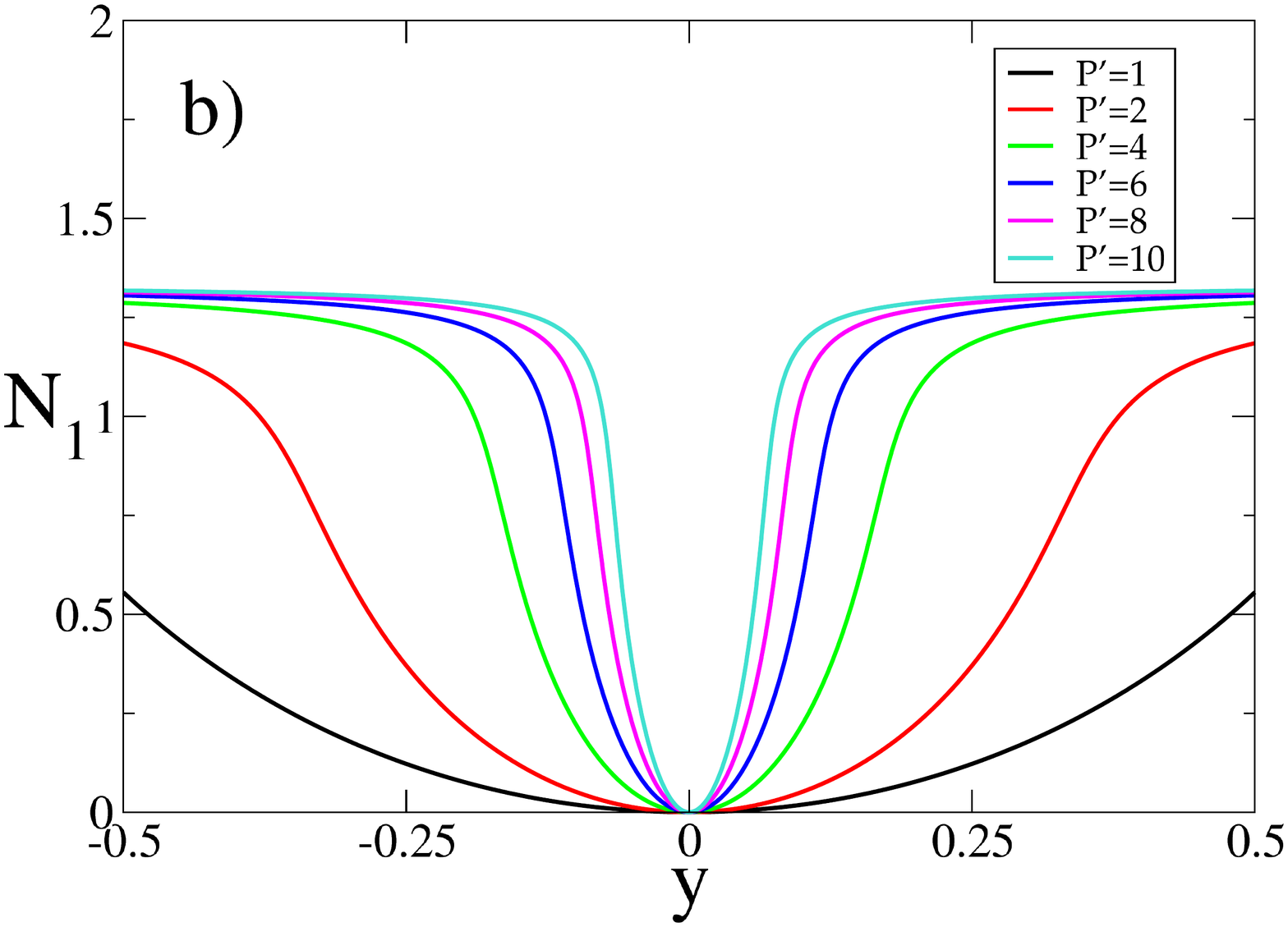}
\caption{(a) Base state profile of shear rate as a function of position across the channel in the Rolie-Poly model for $\beta=0.5,\eta=0.03$. Inset shows the same data for $\gdot>0$ on a log-log scale. (b) Corresponding base state profile of first normal stress as a function of position across the channel.}
\label{fig:RP_base}
\end{figure*}

The linear analysis just described is valid as long as the perturbations remain small. To study the dynamics once the perturbations have grown (in the unstable regime) to attain a finite amplitude, we perform a full nonlinear simulation in the $x-y$ plane, using methods as in Ref.~\cite{fielding2010shear}. We have checked that the early-time dynamics of this full 2D simulation, starting from a 1D base state subject to a small perturbation, agrees with the linear calculation outlined above.

The numerical calculations introduce two further parameters: the numerical timestep $dt$ and the number of numerical gridpoints $N_y$ (and $N_x$ in the fully 2D simulations). We give values for these in figure captions below, but have also checked that the results are robust to further decreases in $dt$ and increases in $N_y, N_x$.

\section{Results}
\label{sec:results}

\subsection{Rolie-Poly model}
\subsubsection{0D flow curves}

We start by discussing the 0D flow curves of the Rolie-Pole model, computed within the assumption of a homogeneous shear flow. (An example would be homogeneous shear between flat parallel plates that translate relative to each other, known as planar Couette flow.) The shear stress as a function of shear rate, $\Sigma(\gdot)$, is plotted in Fig.~\ref{fig:RP_curves}a). For strain rates roughly in the range $\gdot=1-10$, this shows a regime of strong shear thinning in which the stress shows a quasi-plateau at $\Sigma^*\approx 0.75$, in which the stress rises only slowly with strain rate. In contrast, the first normal stress $N_1(\gdot)$ rises steadily to an eventual asymptote as $\gdot\to\infty$.

\subsubsection{1D base state}

We now consider the way in which the 0D flow curves just discussed, computed for a homogeneous shear flow, relate to the 1D base state profiles in a pressure driven channel flow, in which the shear rate and shear stress now vary across the channel. In many places the discussion in this section will refer to the right half of the channel, $y>0$, for definiteness. Corresponding statements apply to the left half, after taking into account the antisymmetry of $\gdot(y)$ and symmetry of $N_1(y)$ about the channel centreline, $y=0$.

We start by noting that force balance requires the total shear stress to vary across the channel as $\Sigma=\dP y$, for a 1D flow. Knowledge of the imposed pressure gradient $\dP$ therefore immediately gives the shear stress as a function of the coordinate $y$ across the channel: starting from the middle of the channel at $y=0$ and moving out towards the right hand wall at $y=L_y/2=1/2$, the shear stress $\Sigma$ that forms the vertical axis of Fig.~\ref{fig:RP_curves}a) increases linearly from $\Sigma=0$ at $y=0$ to $\Sigma=\dP/2$ at $y=1/2$. The corresponding shear rate $\gdot(y)$ at any $y$ can therefore be read off as the value of $\gdot$ in Fig.~\ref{fig:RP_curves}a) corresponding to the given $\Sigma$, at that $y$. In other words, inverting the flow curve $\Sigma(\gdot)$ of Fig.~\ref{fig:RP_curves}a) to give $\gdot=\gdot(\Sigma=\dP y)$ immediately gives the base state shear rate profile $\gdot(y,\dP)$, for any $\dP$. A family of such curves for several values of $\dP$ is shown in Fig. ~\ref{fig:RP_base}a). 

In the same way, combining the shear and normal stress flow curves $\Sigma(\gdot)$ and $N_1(\gdot)$ into a parametric plot $N_1(\Sigma)$, as shown in the inset to Fig.~\ref{fig:RP_curves}b), immediately gives the normal stress $N_1(\Sigma=\dP y)$ as a function of position across the channel. A family of these base state normal stress curves $N_1(y)$ for several values of $\dP$ is shown in Fig.~\ref{fig:RP_base}b). A key feature in each curve (apart from at the lowest value of $\dP$) is the existence of a fairly well localised region for which $N_1(y)$ increases rapidly with $y$. This stems directly from the strong shear thinning seen in Fig.~\ref{fig:RP_curves}a). As the shear stress increases with $y$ across the channel, one will at the position $y^*=\Sigma^*/\dP$ corresponding to the stress quasi-plateau at $\Sigma = \Sigma^*\approx 0.75$ in Fig.~\ref{fig:RP_curves} reach the shear thinning regime in which the shear rate increases strongly with $\Sigma$, and so with $y$. Correspondingly, the normal stress $N_1(\gdot)$ also increases strongly with $y$. This steep rise is only seen for pressure drops above about $\dP\approx 1.5$, however: for lower pressures, the stress $\Sigma=\dP y$ does not attain the shear thinning regime $\Sigma^*\approx 0.75$ before the wall is reached. In this way, the stress at the wall, $\Sigma_{\rm wall}=\dP/2$, determines how far up the axis of Fig.~\ref{fig:RP_curves}a) is explored.

The discussion of the previous paragraph can be expressed mathematically as:
\beqna
\frac{dN_1}{dy}=\frac{d\Sigma}{dy}\frac{dN_1}{d\Sigma}=\dP\frac{dN_1}{d\Sigma}=\dP\frac{dN_1}{d\gdot}/\frac{d\Sigma}{d\gdot}.
\label{eqn:dN1}
\eeqna
From this, we indeed see that the regime of large $dN_1/dy$ in the base state profile $N_1(y)$ in Fig.~\ref{fig:RP_base}b) corresponds to that of strong shear thinning (small $d\Sigma/d\gdot$) in the flow curve of Fig.~\ref{fig:RP_curves}a), for any pressure drop large enough that this shear thinning regime is indeed explored, within the window of stresses $-\dP/2<\Sigma<\dP/2$ inside the channel.

\subsubsection{Instability to 2D perturbations}

\begin{figure}[!t]
\vspace{-0.75cm}
\includegraphics[width=0.475\textwidth]{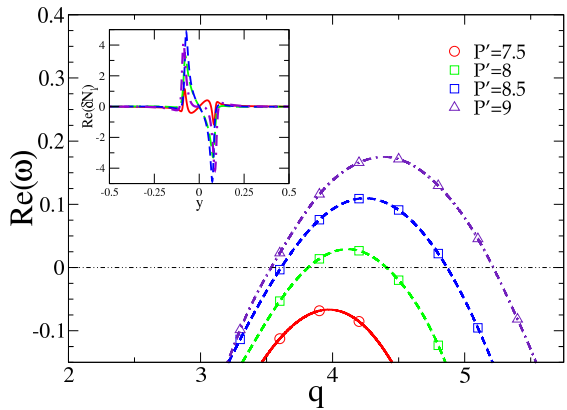}
\caption{Dispersion relation of growth rate as a function of wavevector for a range of pressure drops in the Rolie-Poly model for a CCR parameter $\beta=0.5$ and solvent viscosity $\eta=0.03$. Numerical grid and timestep $N_y=1024$ and $dt=0.0001$.  
Inset: real part of the normalised eigenfunction in the first normal stress difference, $\delta N_1$, at wavevector $q=4.5$.}
\label{fig:LinearBehaviourRP}
\end{figure}

We now report the results of our linear stability analysis for the dynamics of 2D perturbations to the 1D base states just discussed. Recall that we defined above $\omega_q$ to be the eigenvalue with the largest real part, at any wavevector $q$.  The dispersion relation  $\Re\omega_q(q)$ is shown in Fig.~\ref{fig:LinearBehaviourRP}. A value $\Re\omega_q>0$ signifies that the 1D base state is unstable to the growth of a 2D perturbation $\propto \exp(iqx)$. For the pair  of parameter values $\beta,\eta$ in Fig.~\ref{fig:LinearBehaviourRP}, we indeed find instability for imposed pressure drops exceeding a critical threshold value, $\dP>\dPs(\beta,\eta) \approx 7.8$.

The corresponding first normal stress part of the eigenfunction, $\delta N_1(y)$  is shown in the inset of Fig.~\ref{fig:LinearBehaviourRP}, for a wavevector $q=4.5$. We have checked that peaks of the eigenfunction are in essentially the same location as the derivative of the base state, $dN_1/dy$, and with the same properties of symmetry as $y\to -y$. This indicates that the mode of instability corresponds to a displacement of the base state. This displacement is furthermore in the same direction in both halves of the channel, corresponding to a `sinuous' mode. (A `varicose' mode would instead correspond to equal and opposite displacements in the two halves of the channel. Note that `sinuous' vs `varicose' is often defined in terms of the symmetry of the perturbation to the streamfunction, but we have checked that this accords with that of the perturbation to $N_1$.) This will be substantiated by the results of our nonlinear simulations in Fig.~\ref{fig:nonlinear} below. In closely resembling the derivative of the base state, the eigenfunction appears strongly localised in the region in which the first normal stress $N_1(y)$ changes rapidly with $y$.

\begin{figure}[!t]
\vspace{-0.5cm}
\includegraphics[width=0.475\textwidth]{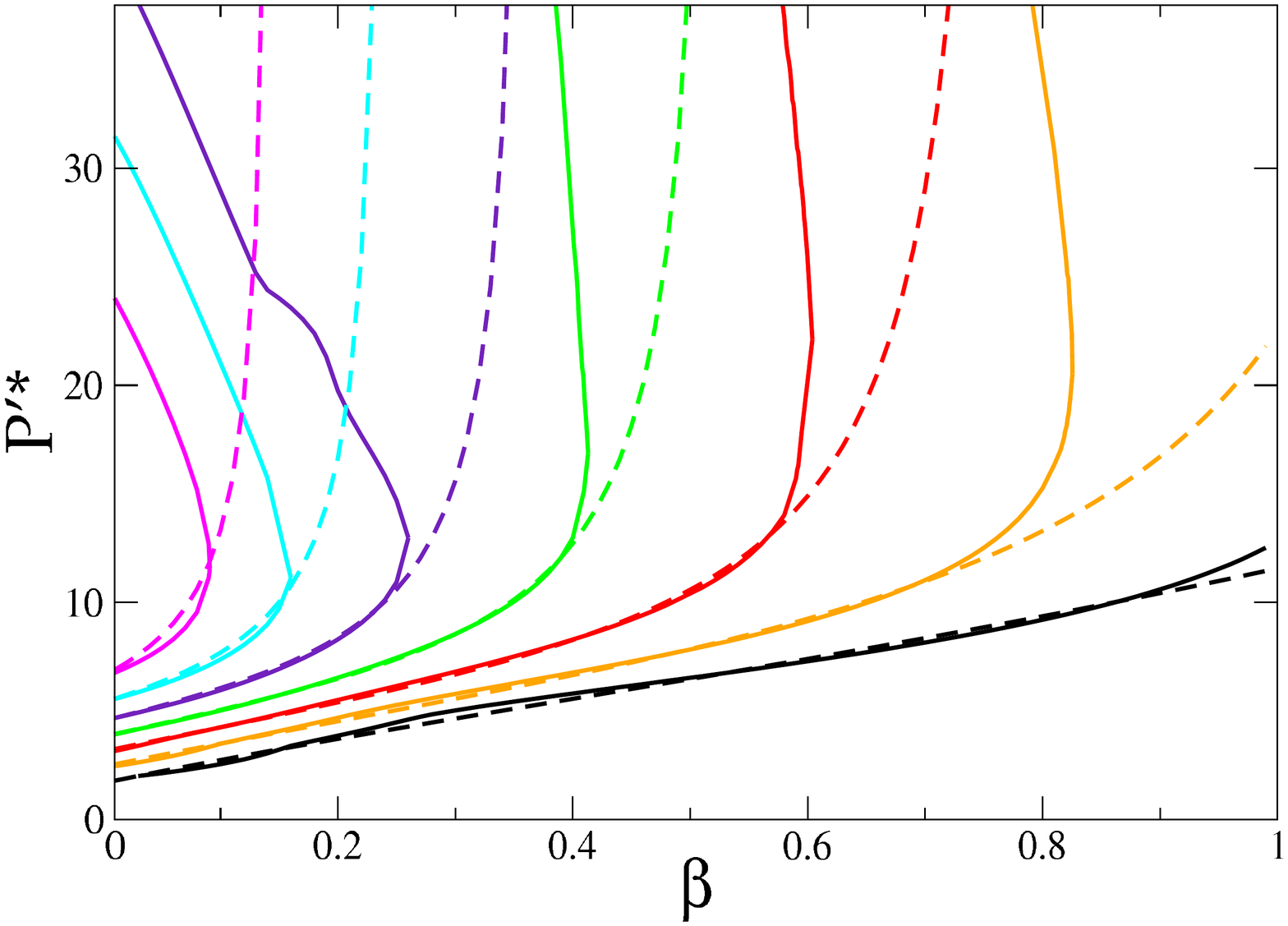}
\caption{Solid lines: curves of neutral stability $\dPs(\beta,\eta)$ as a function of the CCR parameter $\beta$ in the Rolie-Poly model for several values of the solvent viscosity $\eta=\textcolor{black}{0.025}, \textcolor{orange}{0.03}, \textcolor{red}{0.035}, \textcolor{green}{0.04}, \textcolor{indigo}{0.045}, \textcolor{cyan}{0.05}, \textcolor{magenta}{0.055}$ (curves from right to left). The flow is stable for low $\dP$ and (for all values of $\eta$ shown apart from $0.025$) also at high $\dP$, with a window of instability in between. Dashed lines: fits using Eqns.~\ref{eqn:criterion} and~\ref{eqn:h} (which we note does not capture the regime of re-entrant stability). 
\label{fig:RP_stability_limit}} 
\end{figure}

For any pair of parameter values $\beta,\eta$, there either exists a critical value $\dPs(\beta,\eta)$ of the pressure drop above which instability first arises, as in Fig.~\ref{fig:LinearBehaviourRP}, or the flow is stable for all the values of imposed $\dP$ that we have explored. 
By performing linear stability calculations over the full range of values of the convective constraint release parameter, $0<\beta<1$, for several values of the solvent viscosity $\eta$, we find the family of curves of neutral stability $\dPs(\beta,\eta)$ shown by the solid curves in Fig.~\ref{fig:RP_stability_limit}. For any fixed $\beta$, the flow is always stable at low $\dP$. (This is to be expected: the flow becomes Newtonian in this limit.) For the lowest value of solvent viscosity shown, $\eta=0.025$, increasing $\dP$ gives instability onset at a critical $\dPs(\beta,\eta=0.025)$ for all values of $\beta$.  For larger values of $\eta$, there exists at low $\beta$ a window of pressure drops at which instability arises, with re-entrant stability at very high $\dP$; for larger $\beta$, the flow is stable at all $\dP$.

The kinks apparent in some of the solid curves in Fig.~\ref{fig:RP_stability_limit} arise not from numerical difficulties, but because the most unstable mode switches at the location of these kinks, with a corresponding switch of the dominant wavevector $q^*$. 

\subsubsection{Criterion for instability onset}
\label{sec:criterion}

Within the Rolie-Poly model, we have seen that an initially 1D base state flow is linearly unstable to the onset of 2D perturbations over a wide range of pressure drops and model parameter values. Our aim now is to uncover a criterion that relates the pressure drop at which the instability first sets in (the lower boundaries in Fig.~\ref{fig:RP_stability_limit}) to the properties of the underlying 1D base state, and thence to the properties of the 0D flow curves, thereby providing experimentalists with a practical guide for when any given pressure driven flow should become unstable.

\begin{figure}[!t]
\includegraphics[width=0.5\textwidth]{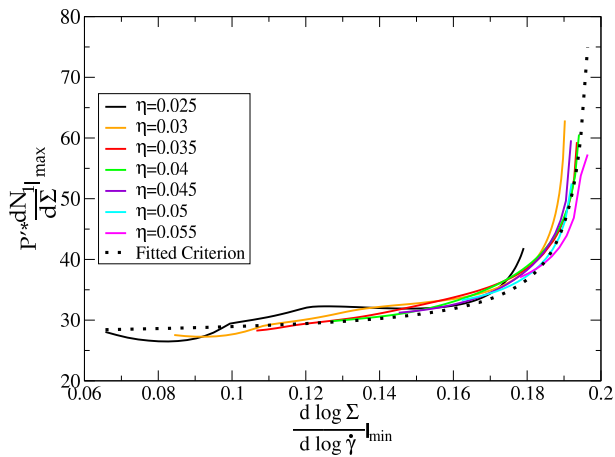}
\caption{The left hand side $y=\dPs dN_1/d\Sigma|_{\rm max}$ of inequality~\ref{eqn:criterion} plotted as a function of the argument $x=d\log\Sigma/d\log\gdot|_{\rm min}$ of $h$ on the right hand side of ~\ref{eqn:criterion}. Dashed line: function $y=h(x)$ of Eqn.~\ref{eqn:h} with $\alpha_0=25.81$, $\alpha_1=0.332$ and $\alpha_2=0.205$. }
\label{fig:RPnormal}
\end{figure}

As noted above, the eigenfunction associated with the instability is strongly localised in the region where $N_1(y)$ changes steeply as a function of position across the channel. Accordingly, we postulate that a high value of $dN_1/dy$, maximised across the channel coordinate $y$, tends to predispose the system to instability. Let us denote this by $dN_1/dy|_{\rm max}$. 

We further recall from Eqn.~\ref{eqn:dN1} that $dN_1/dy$ increases with imposed pressure drop $\dP$. 
Indeed, even a non-shear thinning model such as Oldroyd B would have values of $dN_1/dy$ that increase indefinitely with $dP'$ in a channel flow. We know, however, that channel flow of the Oldroyd B model is linearly stable to 2D perturbations~\cite{wilson1999structure,ho1977stability}. Accordingly, a high value of $dN_1/dy$ alone cannot be a sufficient condition for linear instability. We therefore additionally postulate the presence of shear thinning to be a necessary condition, such that the region of high $dN_1/dy$ is focused into some localised window across the channel, as discussed above for the curves in Fig.~\ref{fig:RP_base}b). We take as a measure of shear thinning the logarithmic slope of the flow curve, minimised across the channel. For any pressure drop exceeding $2\Sigma^*$, where $\Sigma^*$ is the stress at the quasi-plateau in Fig.~\ref{fig:RP_curves}a), this simply equals the value of $d\log\Sigma/d\log\gdot$ minimised on the flow curve, which we accordingly denote $d\log\Sigma/d\log\gdot|_{\rm min}$. We assume finally that the condition for instability is 
\be
\frac{dN_1}{dy}|_{\rm max}>h\left(\frac{d\log\Sigma}{d\log\gdot}|_{\rm min}\right),
\ee
where $h$ is some as yet unknown increasing adimensional function of its argument. (For the inequality to be dimensionally consistent then of course requires a prefactor $G/L_y$ on the right hand side, which is however equal to one in our chosen units.)

\begin{figure}[!t]
\includegraphics[width=0.475\textwidth]{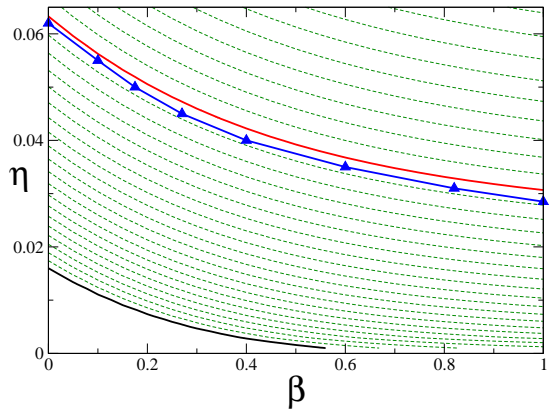}
\caption{Dotted lines: contours of minimum logarithmic slope of the flow curve, $n=d\log\Sigma/d\log\gdot|_{\rm min}$. The lowest (black) contour has $n=0.0$ (below which shear banding occurs), with $n$ increasing in increments of $0.01$ in contours upward. The contour $0.20$ is shown as a solid red line. The triangles show numerical data for the maximum solvent viscosity $\eta$ that admits instability at any value of the CCR parameter, $\beta$, consistent with a contour value $n\approx 0.205$.}
\label{fig:contoursRP}
\end{figure}

\begin{figure*}[t!]
\includegraphics[width=0.475\textwidth]{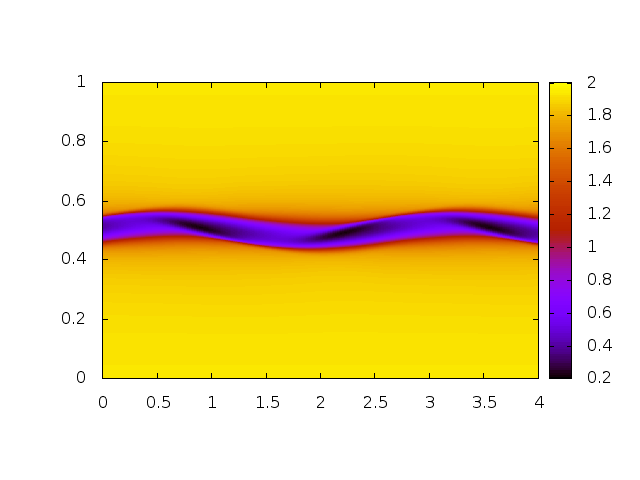}
\includegraphics[width=0.475\textwidth]{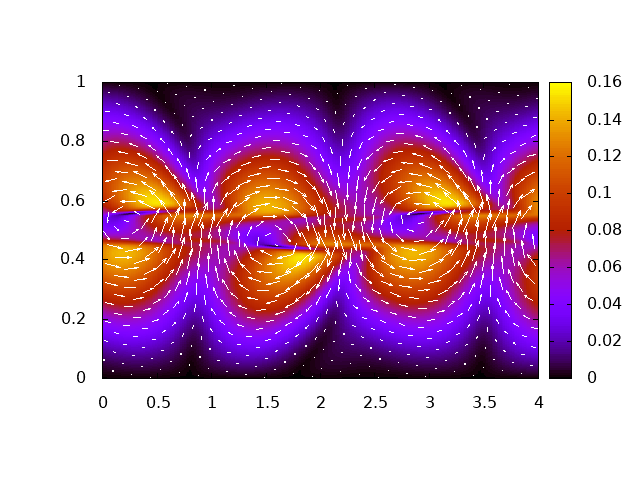}
\caption{Non-linear flow states for the Rolie-Poly model  for parameter values $\beta=0.0, \eta=0.03, \dP=15.0, L_x=16.0$. Numeric grid: $N_x=512$, $N_y=1024$ and timestep $dt=0.0003$.
Left: the component $\sigma_{xx}$ of the polymeric stress tensor. Right: magnitude of the perturbation of the velocity in the final 2D state, compared with the initial 1D base state. The arrows show the direction of this velocity perturbation. Only $1/4$ of the simulation box is shown in the $x$ direction.}
\label{fig:nonlinear}
\end{figure*}

Recalling that $dN_1/dy=\dP dN_1/d\Sigma$, this  condition for instability can equivalently be rewritten as a criterion on the value of the imposed pressure:
\be
\dPs\frac{dN_1}{d\Sigma}|_{\rm max}>h\left(\frac{d\log\Sigma}{d\log\gdot}|_{\rm min}\right),
\label{eqn:criterion}
\ee
in which $dN_1/d\Sigma|_{\rm max}$ is the maximum value of $dN_1/d\Sigma$ at any point across the channel $y$. For pressure drops such that the shear stress at the wall exceeds the quasi-plateau value in Fig.~\ref{fig:RP_curves}a), this simply corresponds to $dN_1/d\Sigma$ maximised over the parametric curves of $N_1(\Sigma)$ in the inset to Fig.~\ref{fig:RP_curves}b). We have confirmed, further, that this corresponds closely to the value of $dN_1/d\Sigma$ as calculated at the shear rate corresponding to the point of minimum logarithmic slope of the constitutive curve. Accordingly, we now use $dN_1/d\Sigma|_{\rm max}$  to denote this value.

To test this criterion, we plot in Fig.~\ref{fig:RPnormal} the left hand side of~\ref{eqn:criterion} as a function of the argument of $h$ on the right hand side. To do so, for each parameter pairing $\beta,\eta$ explored in Fig.~\ref{fig:RP_stability_limit} we read off the critical pressure drop from that Fig.~\ref{fig:RP_stability_limit}, and compute $d\log\Sigma/d\log\gdot$ minimised over the flow curve of Fig.~\ref{fig:RP_curves}a) for that $\beta,\eta$, together with $dN_1/d\Sigma|_{\rm max}$ as just defined (corresponding closely to the value of this quantity maximised over the parametric normal stress curve inset in Fig.~\ref{fig:RP_curves}b).  The results are collected in Fig.~\ref{fig:RPnormal} into a set of curves, with each curve containing all the data for a single value of $\beta$. We obtain reasonable data collapse onto the function
\be
h(x)=\alpha_0+\alpha_1/(\alpha_2-x),
\label{eqn:h}
\ee
with the same set of fitting parameters $\alpha_0=25.81$, $\alpha_1=0.332$ and $\alpha_2=0.205$ for all $\beta,\eta$.

Now that this function $h(x)$ is known, we can finally reconstruct the critical pressure drop for any pair of parameter values $\beta,\eta$ by computing $h(x)/y$ with $x=d\log\Sigma/d\log\gdot_{\rm min}$ (for that $\beta,\eta$) and $y=dN_1/d\Sigma|_{\rm max}$ (for that $\beta,\eta$). This gives the dashed lines in Fig.~\ref{fig:RP_stability_limit}. These indeed fit the numerical data quite well at low values of the imposed pressure, thereby capturing the initial onset of instability. The fits however fail to capture the regime of re-entrant stability, and so depart from the numerical data around the `nose' in that data.

The form of the function $h$ reveals that instability can never be attained for a value of the logarithmic derivative of the flow curve, minimised across the flow curve,  $d\log\Sigma/d\log\gdot|_{\rm min}$, that exceeds a value roughly equal to $0.2$. This sets a basic condition on how flat the logarithmically plotted flow curve must be to admit instability: were the flow curve to be approximated by a power law $\Sigma\sim\gdot^n$ over the quasi-plateau region, instability requires $n<0.21$, in the Rolie-Poly model. This is broadly consistent with instability having been observed for the shear thinning exponents of $n=0.21$ and $n=0.19$ in the experimental studies of Refs.~\cite{Bodiguel2015} and~\cite{Poole2016} respectively, but apparently at odds with instability having been observed for $n=0.30$ in the experiments of Ref.~\cite{Picaut2017}.

Indeed, in Fig.~\ref{fig:contoursRP} we represent the value of $d\log\Sigma/d\log\gdot|_{\rm min}$ by dotted contour lines in the plane of the model parameters $\eta$ and $\beta$. We also show by a cross the value of solvent viscosity below which we find instability, at any pressure drop, for any value of $\beta$. The regime of instability is indeed found to be that for which the minimum logarithmic derivative is less than about $0.21$. 

The thick black line in Fig.~\ref{fig:contoursRP} shows the contour $n=0$, below which the constitutive curve of stress against strain rate is non-monotonic, admitting shear banding. Previous studies~\cite{fielding2005linear,fielding2010shear,fielding2006nonlinear} in the {\em Johnson-Segalman} model showed an initially 1D shear banded state to be linearly unstable to the onset of 2D perturbations in which the interface between the bands destabilises. While we do not explore this region in detail here, we have confirmed that the interface is indeed unstable in the Rolie-Poly model, increasing the generality of the previous finding in the Johnson-Segalman model. (To extend our study into the banding regime, diffusive terms were added to the right hand side of the Rolie-Poly constitutive equation~\cite{lu2000effects}.) In this way, the instability of a strongly shear thinning -- but not shear banding --  fluid that is the main focus of this work can potentially be understood as an instability of the quasi-interface in the region of steeply increasing $N_1(y)$ in Fig.~\ref{fig:RP_base}b).The addition of a stress diffusion term outside the banding regime would have the effect of reducing the steepness of variation of $N_1(y)$ within the channel, thereby potentially having a slightly stabilising effect on the overall flow~\cite{wilson1997short}.

\begin{figure*}[!t]
\includegraphics[width=0.4\textwidth]{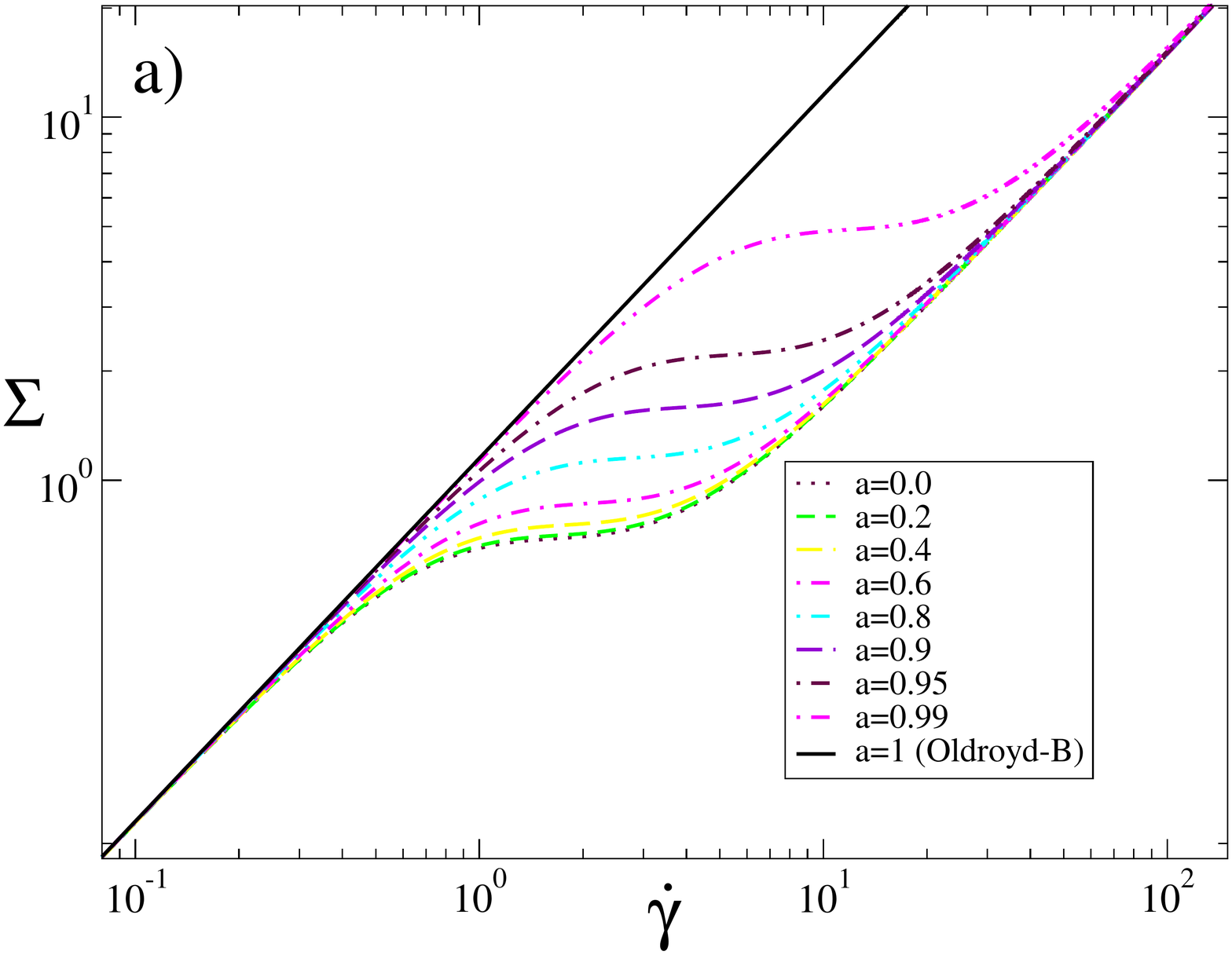}
\includegraphics[width=0.4\textwidth]{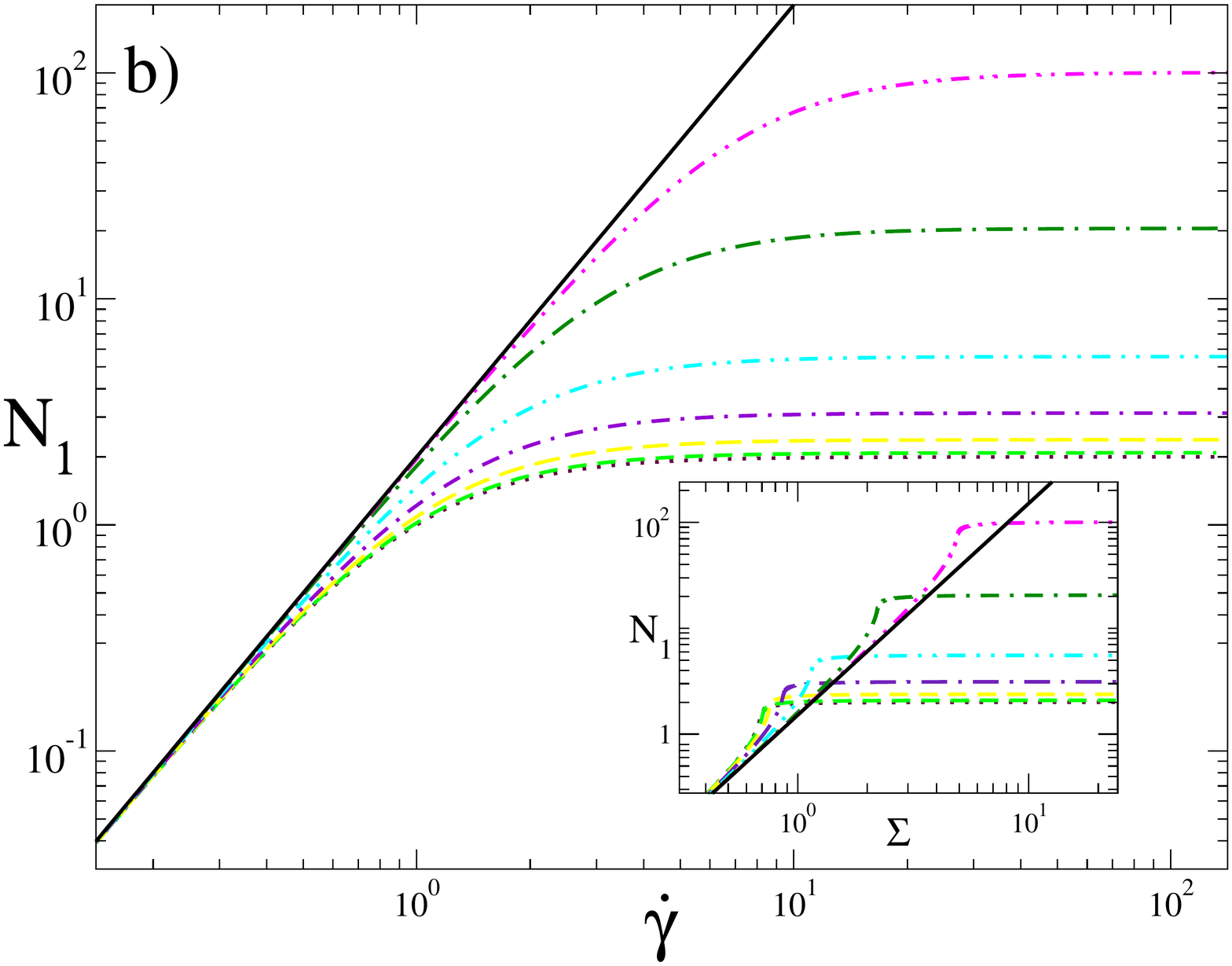}
\caption{(a) Flow curves of shear stress as a function of shear rate for states of homogeneous shear flow in the Johnson-Segalman model, for a solvent viscosity $\eta=0.15$ and several values of the slip parameter $a$. (b) The corresponding curves of normal stress as a function of shear rate, and (inset) parametrically plotted as a function of shear stress.}
\label{fig:JS_flow_curves}
\end{figure*}

\subsubsection{Non-linear dynamics}
\label{sec:nonlinear}

The linear analysis just described is valid in the regime where the 2D perturbations to the 1D base state remain small. To study the dynamics once the perturbations have grown (in the unstable regime) to attain a finite amplitude, we performed fully nonlinear 2D simulations.  Sample snapshots of the ultimate state are shown in Fig.~\ref{fig:nonlinear}. The left panel confirms the basic mode of instability to be one of displacement of  a quasi-interface between regions of low and high first normal stress. The right panel shows the accompanying difference between the velocity fields of the final 2D and initial 1D velocity fields, revealing a system-spanning vortex (which is modest in amplitude compared with the velocities in the underlying base state). We note that the amplitude of velocity perturbation seen here in our nonlinear simulations is smaller compared with the base flow, than in the experiments of Ref.~\cite{Bodiguel2015, Poole2016}. 

\subsection{Johnson-Segalman model}
\subsubsection{0D flow curves}

The flow curves of the Johnson-Segalman model, describing states of homogeneous shear flow, are shown in Fig.~\ref{fig:JS_flow_curves}. These have the same basic form as for the Rolie-Poly model in Fig.~\ref{fig:RP_curves}, with a stress quasi-plateau associated with a regime of strong shear thinning. An important difference between the two models, however, is that whereas the stress $\Sigma^*$ at the quasi-plateau in the Rolie-Poly model is more or less independent of the CCR parameter, $\beta$, in the Johnson-Segalman model it depends strongly on the slip parameter $a$, scaling as:
\be
\Sigma^*\sim\dfrac{G}{\sqrt{1-a^2}}.
\ee
This diverges in the limit $a\to 1$, in which the Johnson-Segalman model becomes the Oldroyd B model.

\begin{figure*}[!t]
\includegraphics[width=0.475\textwidth]{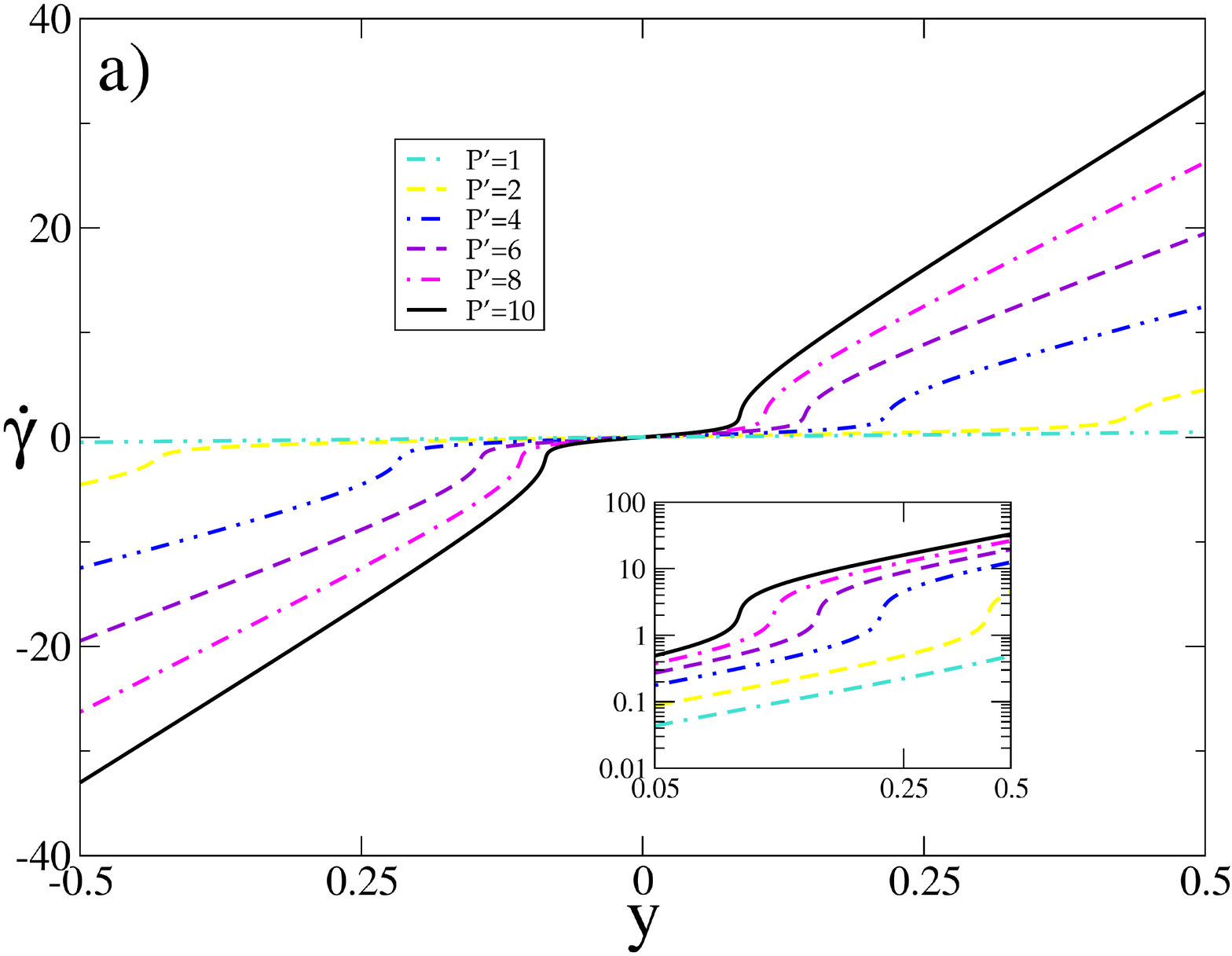}
\includegraphics[width=0.475\textwidth]{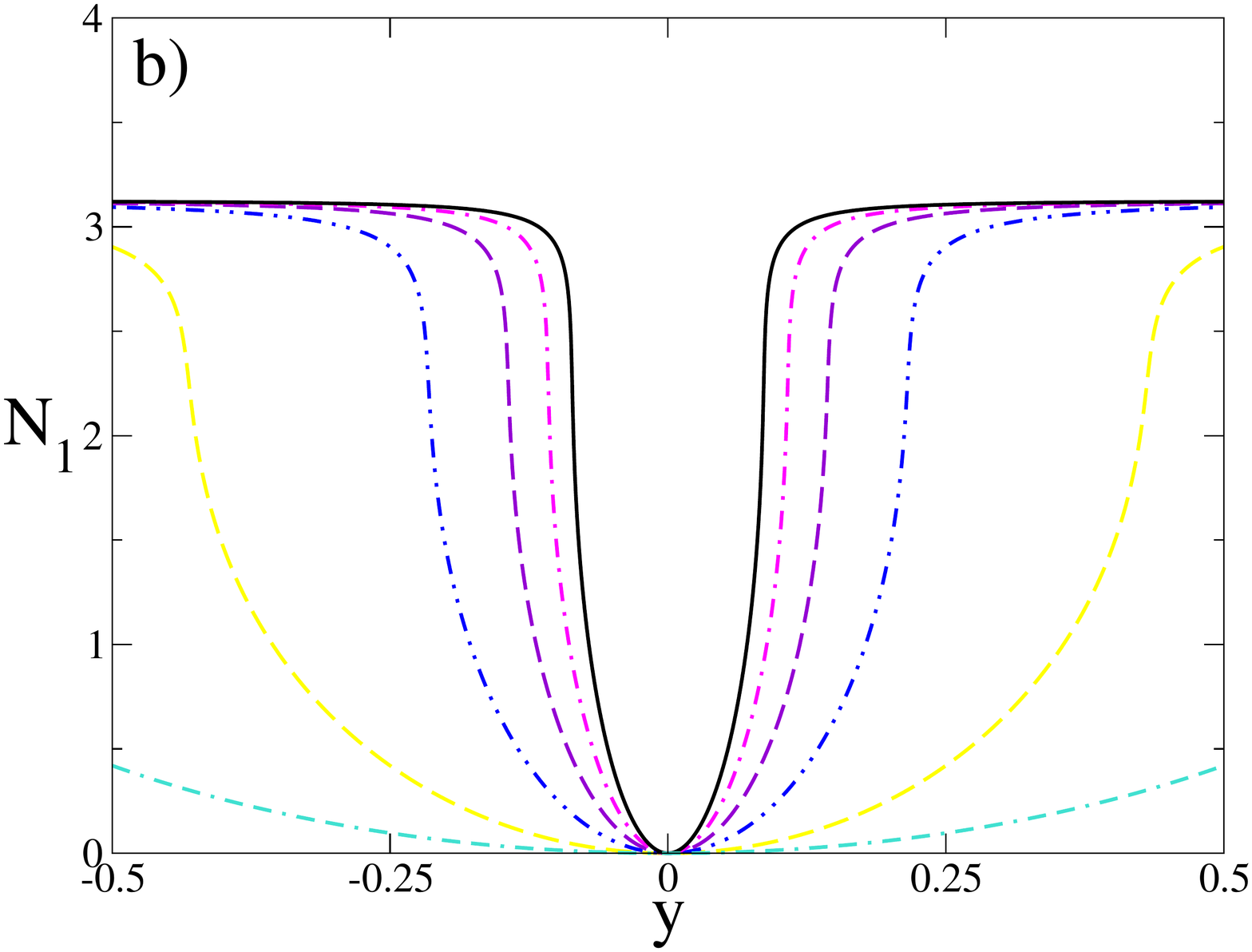}
\caption{(a) Base state profile of shear rate as a function of position across the channel in the Johnson-Segalman model for $a=0.6,\eta=0.15$, $N_y=2048$. Inset shows the same data for $\gdot>0$ on a log-log scale. (b) Corresponding base state profile of first normal stress as a function of position across the channel.}
\label{fig:JS_base}
\end{figure*}

A second difference between the Rolie-Poly and Johnson-Segalman models is that the minimum logarithmic slope of the flow curve, $d\log\Sigma/d\log\gdot|_{\rm min}$, is independent of the slip parameter $a$ in the Johnson-Segalman model, whereas  it depends strongly on the CCR parameter $\beta$ in the Rolie-Poly model. (Recall Fig.~\ref{fig:contoursRP}.)

\subsubsection{1D base states}

The 1D base states $\gdot(y)$ and $N_1(y)$ of the Johnson-Segalman model are shown in Fig.~\ref{fig:JS_base}. These can be related to the 0D flow curves of Fig.~\ref{fig:JS_flow_curves} in the same way as for the Rolie-Poly model. In particular, the regime of strong shear thinning in the flow curve $\Sigma(\gdot)$ gives rise to the region of rapid variation of shear rate $\gdot(y)$ and first normal stress $N_1(y)$ as a function of the position $y$ across the channel.

\subsubsection{Instability to 2D perturbations}

\begin{figure}[!t]
\includegraphics[width=0.49\textwidth]{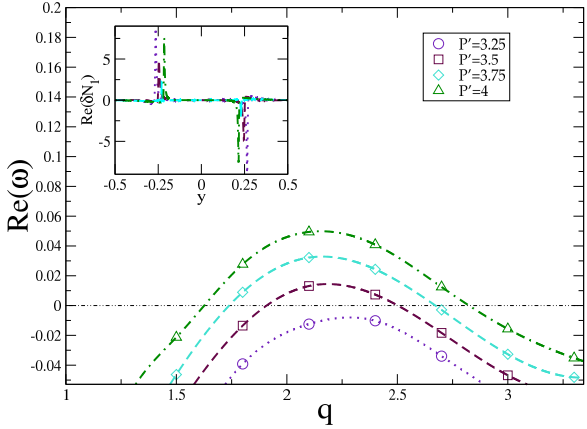}
\caption{Dispersion relation of growth rate as a function of wavevector for a range of pressure drops in the Johnson-Segalman model for a slip parameter $a=0.6$ and solvent viscosity $\eta=0.15$. Numerical grid $N_y=2048$ and timestep $Dt=0.0001$.
Inset: real part of the normalised eigenfunction in the first normal stress difference, $\delta N_1$, at wavevector $q=2.7$.   }
\label{fig:JS_dispersion}
\end{figure}

We now consider whether the 1D base states just discussed are stable or unstable to the growth of 2D perturbations, for any set of model parameter values $a,\eta$, channel length $L_x$ and imposed pressure drop.
For a particular pair of values of $a,\eta$, we show in Fig.~\ref{fig:JS_dispersion} the dispersion relation of the real part of the most unstable eigenvalue as a function of wavevector $q$ (which recall must be quantised as $n\pi/L_x$ for any channel of finite length $L_x$.) As in the Rolie-Poly model, we find onset of instability above a critical pressure drop $\dPs(\eta,a)$. The dispersion relations have very similar form to those in the Rolie-Poly model. (Recall Fig.~\ref{fig:LinearBehaviourRP}.) So too does the associated first normal stress part of the eigenfunction $\delta N_1(y)$, as shown in the inset to Fig.~\ref{fig:JS_dispersion}. Indeed, it is heavily localised in the region of steep gradient in $N_1(y)$ of the underlying base state. The instability  therefore again appears to correspond to a displacement of the quasi-interface formed from the region of steep variation in $N_1(y)$ across the channel in the underlying base state. This will be confirmed by the results of our nonlinear simulations below. 

This is consistent with earlier studies of the Johnson-Segalman model in the shear banding regime, $\eta<0.125$, in which a 1D shear banding flow was shown to be unstable to 2D undulations along the interface between the bands~\cite{fielding2005linear,fielding2010shear,fielding2006nonlinear}, as noted above. However, we do not consider this shear banding regime here, confining ourselves instead to values of the solvent viscosity $\eta>0.125$ for which the constitutive curve is monotonic.

By performing linear stability calculations over a range of values of the slip parameter, $0<a<1$, for several values of the solvent viscosity $\eta$, we find a family of curves of neutral stability $\dPs(a,\eta)$ in Fig.~\ref{fig:JS_onset}. For any fixed $a$, the flow is always stable at low $\dP$. (This is to be expected: the flow becomes Newtonian in this limit.) For the lowest four values of solvent viscosity explored in Fig.~\ref{fig:JS_onset}, we find a critical value of pressure drop $\dPs$ above which instability arises, at all values of $a$. For the highest value of viscosity explored, there is a restricted window of values of $a$, above about $0.7$, that admit instability.

We also explored much higher values of imposed pressure drops, up to $\dPs=60.0$, for the single value of solvent viscosity $\eta=0.16$. The results (not shown) reveal a regime of re-entrant stability above a pressure drop of, for example, about $35$ for $a=0.0$, and about $55$ for $a=0.8$. We do not explore this further here: such pressure drops may in any case be unattainable in practice.

\subsubsection{Criterion for onset of instability}

We now seek to determine whether the criterion~(\ref{eqn:criterion}) developed above in the Rolie-Poly model, relating the pressure drop at which the first instability sets in to the properties of the underlying 1D base state, and thence to the the 0D flow curves, also applies in the Johnson-Segalman model. Recall that this criterion reads:
\be
\frac{dN_1}{dy}|_{\rm max}=\dPs\frac{dN_1}{d\Sigma}|_{\rm max}>h\left(\frac{d\log\Sigma}{d\log\gdot}|_{\rm min}\right),
\label{eqn:criterionJS}
\ee
with
\be
h(x)=\alpha_0+\alpha_1/(\alpha_2-x),
\label{eqn:hJS}
\ee
in which $d\log\Sigma/d\log\gdot_{\rm min}$ refers (for each $a,\eta$) to the minimum logarithmic derivative of the flow curve in Fig.~\ref{fig:JS_flow_curves}a), and $dN_1/d\Sigma|_{\rm max}$  refers (for each $a,\eta$) to the maximum derivative of the parametrically plotted normal stress curve in the inset to Fig.~\ref{fig:JS_flow_curves}b). (In fact, it refers to the value of this derivative computed at a shear rate corresponding to the point of minimuum logarithmic slope of the flow curve. This coincides very nearly with the maximum value of  $dN_1/d\Sigma$.) 

%

\begin{figure}[!t]
\includegraphics[width=0.475\textwidth]{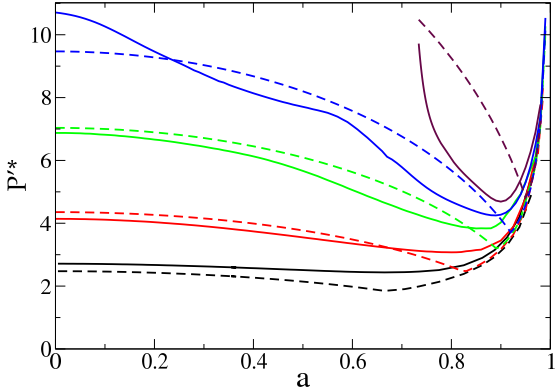}
\caption{Curves of neutral stability $\dPs(a,\eta)$ as a function of the slip parameter $a$ in the Johnson-Segalman model for several values of the solvent viscosity $\eta=$\textcolor{black}{0.14}, \textcolor{red}{0.15}, \textcolor{green}{0.16}, \textcolor{blue}{0.165}, \textcolor{maroon}{0.17} (curves upwards). An initially 1D base state flow is stable below each curve, and unstable above it (until a region of re-entrant stability is reached at much larger pressure drops, not shown).}
\label{fig:JS_onset}
\end{figure}

Plotting the left hand side of~\ref{eqn:criterionJS} as a function of the argument of $h$ on the right hand side, we find a reasonable fit, now with $\alpha_0=60.0,\alpha_1=1.15$ and $\alpha_2=0.11$.  This gives the fitted onset boundaries shown by dashed lines in Fig.~\ref{fig:JS_onset}, for values of $a$ less than around $0.6$. As can be seen, it performs reasonably well for the lowest values of solvent viscosity explored, where the pressure onset values are modest. It performs less well, however, for higher solvent viscosities, for which the pressure onset values are much larger. This could be due to the fact that the quasi-interfacial region of steep $N_1(y)$ becomes close (on the scale of its own thickness) to the centre of the channel at such large pressure drops: a feature not taken into account in our analysis. However, we ignore this complication, on the grounds that such pressure drops may be difficult to obtain in practice in any case.

\begin{figure}
\includegraphics[width=0.475\textwidth]{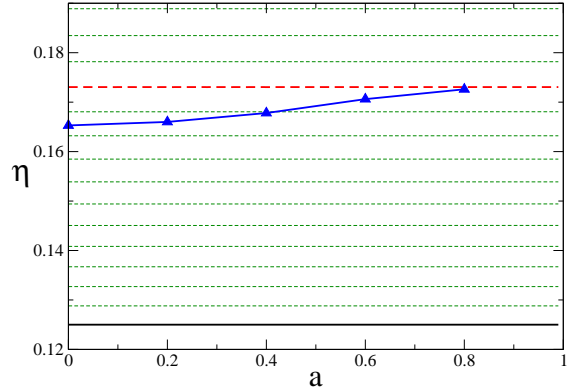}
\caption{Dotted lines: contours of minimum logarithmic slope of the flow curve, $n=d\log(\Sigma)/d\log\gdot|_{\rm min}$. The lowest (black) contour has $n=0$ (below which shear banding occurs), with $n$ increasing in increments of $0.01$ in contours upward. The contour $0.11$ is also shown as a dashed line.  The triangles show numerical data for the maximum solvent viscosity $\eta$ that admits instability at any value of the slip parameter, $a$, showing reasonable agreement with an approximate contour range $n=0.095$ to $n=0.11$.}
\label{fig:contoursJS}
\end{figure}

The value $\alpha_2=0.11$ obtained in the above fitting shows that the value of the logarithmic slope of the flow curve in the quasi-plateau thinning regime should not exceed $n=0.11$ in any fluid for instability to be observed. This sets a basic requirement on how steep the (shallowest part of the) flow curve must be to avoid instability, as explored further in Fig.~\ref{fig:contoursJS}. The value $0.11$ in the Johnson-Segalman model shows that this model predicts, overall, greater stability than the Rolie-Poly model: a fluid has to shear thin more strongly to be predicted unstable by Johnson-Segalman than by Rolie-Poly.

\begin{figure*}[t!]
\includegraphics[width=0.475\textwidth]{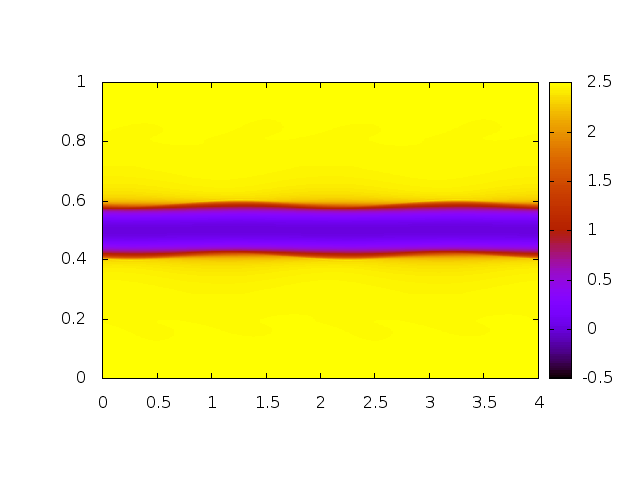}
\includegraphics[width=0.475\textwidth]{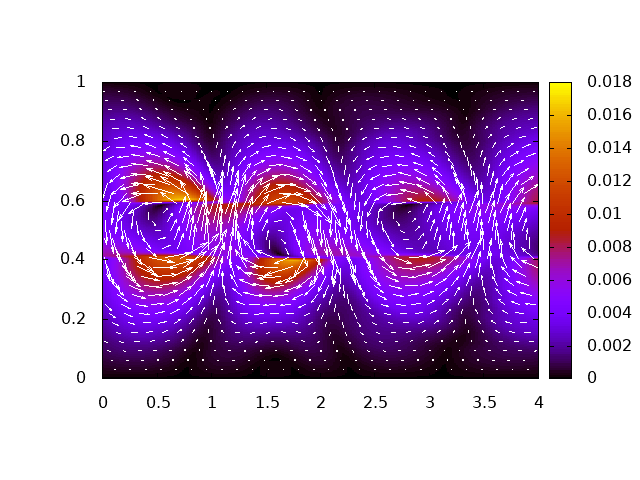}
\caption{Non-linear flow states for the Johnson-Segalman model  for parameter values $a=0.6, \eta=0.16, \dP=10.0, L_x=8.0$. Numeric grid: $N_x=512$, $N_y=1024$ and timestep $Dt=0.0005$.
Left: the component $\sigma_{xx}$ of the polymeric stress tensor. Right: colour scale shows the magnitude of the perturbation of the velocity in the final 2D state, compared with the initial 1D base state. The arrows show the direction of this velocity perturbation. Only half of the simulation box is shown in the $x$ direction.}
\label{fig:nonlinearJS}
\end{figure*}

Whereas the procedure just discussed proved successful in the Rolie-Poly model for all values of the CCR parameter $\beta$, in the Johnson-Segalman model it is restricted to values of the slip parameter $a$ that are not too close to $1.0$. (We used only $0<a<0.6$ in the fitting just described.) For values of $a$ close to $1.0$, the height of the stress quasi-plateau in the flow curves diverges, with the Johnson-Segalman model tending to the Oldroyd B model in the limit $a\to 1$: recall Fig.~\ref{fig:JS_flow_curves}a). This brings an additional requirement for instability: that the imposed pressure drop is large enough that the range of shear stresses within the channel, $0<|\Sigma|<\dP L_y/2$ encompasses the value of the stress quasi-plateau $\Sigma^*\sim G/\sqrt{1-a^2}$. This is necessary to ensure that the region in which $N_1(y)$ varies rapidly with $y$ is indeed present within the channel. Accordingly, we update our criterion to now be a {\em double requirement}: both that  ~\ref{eqn:criterionJS} is satisfied, and that
\be
\dP L_y-2\Sigma^*>0.
\label{eqn:criterionJS1}
\ee
It is this double criterion that is represented by the dashed lines in Fig.~\ref{fig:JS_onset}, with condition~\ref{eqn:criterionJS} being more stringent for lower values of $a$, and criterion~\ref{eqn:criterionJS1} more stringent for higher values of $a$. (In the Rolie-Poly model, we did not need to state the additional condition~\ref{eqn:criterionJS1} explicitly, because it is automatically satisfied in all regimes where ~\ref{eqn:criterion} also holds.)

\subsubsection{Non-linear dynamics}
\label{sec:nonlinearJS}

To study the dynamics once the perturbations have grown to attain a finite amplitude, we performed fully nonlinear 2D simulations.  Sample snapshots of the ultimate state are shown in Fig.~\ref{fig:nonlinearJS}. The left panel confirms the basic mode of instability to be one of displacement of  a quasi-interface between regions of low and high first normal stress, as in the Rolie-Poly model. (Compare Fig.~\ref{fig:nonlinear}.) The right panel shows the accompanying difference between the velocity fields of the final 2D and initial 1D velocity fields, again revealing a system-spanning vortex. 

\subsection{White-Metzner model}

\subsubsection{0D flow curves}

The flow curves of the White-Metzner model can be calculated analytically. In our usual units in which the modulus $G=1$ and the basic timescale $\tau=1$, one has $\Sigma=\gdot|\gdot|^{n-1}+\eta\gdot$ and $N_1=\gdot^{2n}$. An important point to note here is that, although the flow curve does display shear thinning in this model, it does so with a constant logarithmic slope, $d\log\Sigma/d\log\gdot=n$ (at least up to small corrections arising from the solvent viscosity). Accordingly, the region of shear thinning is not focused into a narrow stress window in which the flow curve shows a quasi-plateau, as for the Rolie-Poly model in Fig.~\ref{fig:RP_curves} and the Johnson-Segalman model in Fig.~\ref{fig:JS_flow_curves}.

\subsubsection{1D base states}

\begin{figure}[!t]
\includegraphics[width=0.475\textwidth]{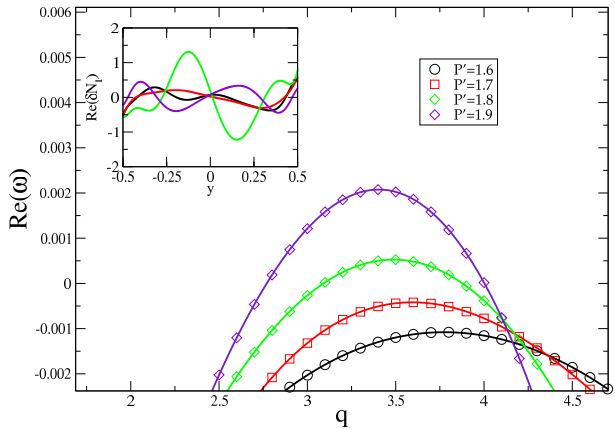}
\caption{Dispersion relation of growth rate as a function of wavevector for a range of pressure drops in the White-Metzner model for a shear thinning index $n=0.2$ and solvent viscosity $\eta=0.005$. Numerical grid  $N_y=2048$ and timestep $dt=0.00005$.
Inset: real part of the normalised eigenfunction in the first normal stress difference, $\delta N_1$, at wavevector $q=3.5$.}
\label{fig:WM_dispersion}
\end{figure}

Combining the form of the White-Metzner model's flow curve $\Sigma=\gdot|\gdot|^{n-1}$ (in which we ignore small corrections in due to the solvent viscosity) with the fact that the shear stress varies as function of position across the channel as $\Sigma=\dP y$, enables us to compute the shear rate as a function of position across the channel in the 1D base state: $\gdot=\sign(y)(\dP |y|)^{1/n}$. Combining this with the form of the normal stress as a function of strain rate, $N_1=\gdot^{2n}$, then gives the normal stress as a function of position across the channel, $N_1=(\dP y)^2$ (as long as $\eta\gdot$ remains small). An important consequence of the absence of any quasi-plateau in the White-Metzner model's flow curve is that there is no quasi-interface across which the first normal stress varies rapidly as a function of position across the channel, as there is for the Rolie-Poly model in Fig.~\ref{fig:RP_base} and for the Johnson-Segalman model in Fig.~\ref{fig:JS_base}. In consequence, we expect the basic physics of any instability to differ from the interfacial mode found for Rolie-Poly and Johnson-Segalman. 

\subsubsection{Instability to 2D perturbations}

Instability has previously been reported in the White-Metzner model by Wilson et al.~\cite{wilson1999instabilityWM,Castillo2017,Wilson2015}. The purpose of our studying the same model again in this work is to examine whether the results can be cast into the same basic format as for the Rolie-Poly and Johnson-Segalman models above, and thereby to try to establish the extent to which any instability of shear thinning channel flow is generic across constitutive models.

We therefore now consider whether the 1D base states just discussed are stable or unstable to the onset of 2D perturbations, for any set of model parameter values $n,\eta$, channel length $L_x$ and imposed pressure drop.
For a particular pair of values of $n,\eta$, we show in Fig.~\ref{fig:WM_dispersion} the dispersion relation of the real part of the most unstable eigenvalue as a function of wavevector $q$ (which recall must be quantised as $n\pi/L_x$ for any channel of finite length $L_x$). As in the Rolie-Poly and Johnson-Segalman models, we find onset of instability above a critical pressure drop $\dPs(\eta,n)$. The dispersion relations have broadly similar form to those for the Rolie-Poly and Johnson-Segalman models in Figs.~\ref{fig:LinearBehaviourRP} and~\ref{fig:JS_dispersion}.

The associated first normal stress part of the eigenfunction, $\delta N_1(y)$, is shown in the inset to Fig.~\ref{fig:WM_dispersion}. A crucial difference between the White-Metzner model and the other two models is that the eigenfunction is diffusely spread across the channel in White-Metzner, rather than being heavily localised, as in Rolie-Poly and Johnson-Segalman. This is consistent with the absence of any steep gradient in $N_1(y)$ in the underlying base state. The instability in White-Metzner therefore does not appear to correspond to a displacement of any quasi-interface formed from a region of steep variation in $N_1(y)$ across the channel in the underlying base state. 

\begin{figure}[!t]
\includegraphics[width=0.475\textwidth]{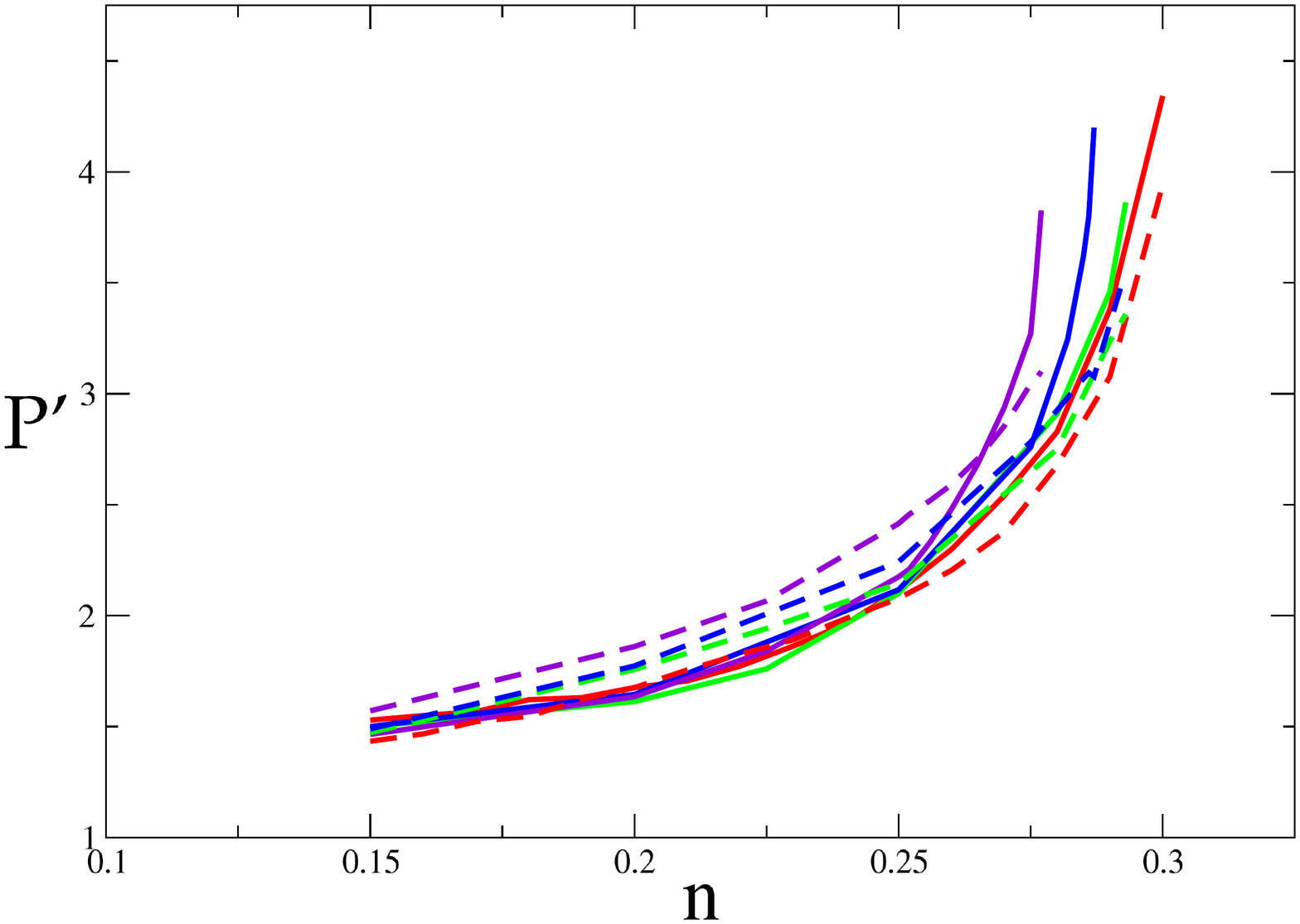}
\caption{Solid lines: curves of neutral stability $\dPs(n,\eta)$ as a function of the shear thinning parameter $n$ in the White-Metzner model for several values of the solvent viscosity solvent values $\eta=\textcolor{red}{0.005}, \textcolor{green}{0.01}, \textcolor{blue}{0.02}, \textcolor{indigo}{0.04}$ in curves upwards. The flow is stable for low $\dP$ and unstable at high $\dP$. (We have not explored higher values of $\dP$ in this model so do not know whether a region of re-entrant stability exists at higher $\dP$ still.) Dashed line: fit using Eqns.~\ref{eqn:criterionWM} and~\ref{eqn:hWM}.}
\label{fig:WM_map1}
\end{figure}

By performing linear stability calculations over a range of values of the shear thinning parameter, $0<n<1$, for several values of the solvent viscosity in the physically relevant regime $\eta<G\tau=1$, we find a family of curves of neutral stability $\dPs(n,\eta)$ in Fig.~\ref{fig:WM_map1}. 

\subsubsection{Criterion for onset of instability}

In the Rolie-Poly and Johnson-Segalman models, we constructed a criterion for the critical pressure drop needed to observe instability.  Recall that this reads:
\be
\frac{dN_1}{dy}|_{\rm max}=\dPs\frac{dN_1}{d\Sigma}|_{\rm max}>h\left(\frac{d\log\Sigma}{d\log\gdot}|_{\rm min}\right),
\label{eqn:criterionWM}
\ee
with
\be
h(x)=\alpha_0+\alpha_1/(\alpha_2-x),
\label{eqn:hWM}
\ee
in which $d\log\Sigma/d\log\gdot|_{\rm min}$ refers (for each $\xi,\eta$) to the minimum logarithmic derivative of the flow curve, and $dN_1/d\Sigma|_{\rm max}$  refers (for each $\xi,\eta$) to the maximum derivative of the parametrically plotted normal stress as a function of shear stress. In the White-Metzner model,  $d\log\Sigma/d\log\gdot=n$ is constant across the flow curve (to within small corrections set by the solvent viscosity). Accordingly, we drop the subscript ``min" from $d\log\Sigma/d\log\gdot|_{\rm min}$. As noted above, this absence of any quasi-plateau in the flow-curve means that there is no strongly localised region of high gradient in $dN_1/dy$ forming a quasi-interface at some location across the channel. Instead, $dN_1/d\Sigma$ is now maximum close to the channel wall, and it is at this location that we accordingly now calculate $dN_1/d\Sigma|_{\rm max}$. This derivative therefore cannot be calculated from a one-off inspection of the flow curve in White-Metnzer, but must involve explicit knowledge of the stress at the wall of the channel. 

\begin{figure}[!t]
\includegraphics[width=0.475\textwidth]{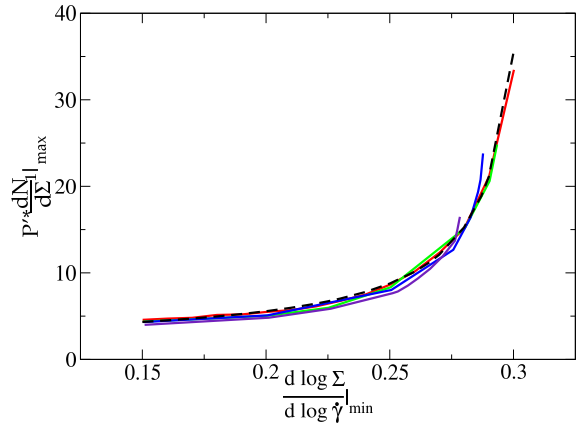}
\caption{The left hand side $y=\dPs dN_1/d\Sigma|_{\rm max}$ of inequality~\ref{eqn:criterionWM} plotted as a function of the argument $x=d\log(\Sigma)/d\log\gdot=n$ of $h$ on the right hand side of~\ref{eqn:criterion}. Dashed line: function $y=h(x)$ of Eqn.~\ref{eqn:h} with $\alpha_0=1.21$, $\alpha_1=0.55$ and $\alpha_2=0.31$.}
\label{fig:WM_criterion}
\end{figure}

\begin{figure}[!t]
\includegraphics[width=0.475\textwidth]{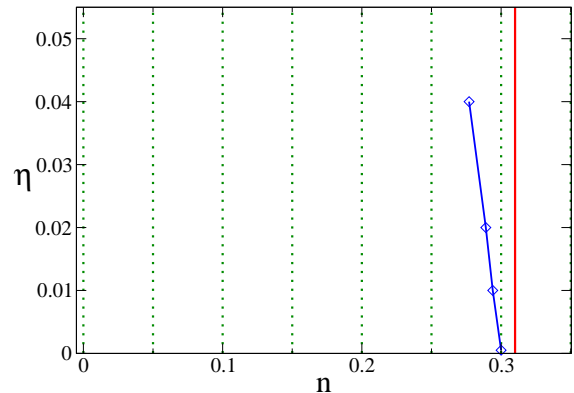}
\caption{Dotted lines: contours of minimum logarithmic slope of the flow curve, $n=[d\log(\Sigma)/d\log\gdot|_{\rm min}]$, increasing in increments of $0.05$ in contours rightward. The contour $n=0.314$ is shown as a solid red line. The diamonds show numerical data for the maximum solvent viscosity $\eta$ that admits instability at any value of the shear thinning parameter $n$, consistent with a contour value in the range $0.277$ to $0.3$.}
\label{fig:WM_map2}
\end{figure}

The criterion  just discussed was built on arguments concerning the existence of a quasi-interface in $N_1(y)$ in the underlying base state of 1D channel flow, in the Rolie-Poly and Johnson-Segalman models. This stemmed in turn from the quasi-plateau in the models' basic flow curves $\Sigma(\gdot)$ for homogeneous shear flow. In view of the absence of any quasi-plateau (and corresponding quasi-interface) in the White-Metzner model, there is no reason, a priori, to expect the criterion still to apply. Nonetheless, we explore whether it does by plotting $dN_1/d\Sigma|_{\rm max}$ as a function of $d\log(\Sigma)/d\log(\gdot)=n$ in Fig.~\ref{fig:WM_criterion}. As can be seen, a good fit to the function $h$ is obtained, with $\alpha_0=1.21,\alpha_1=0.55$ and $\alpha_2=0.31$.  This then enables us to reconstuct a fit to the critical pressure drop as a function of $\eta,n$, as shown by the dashed line in Fig.~\ref{fig:WM_map1}, giving reasonable agreement with the numerical data. Finally, we show in Fig.~\ref{fig:WM_map2} contours of the logarithmic slope of the flow curve (minimised over the flow curve) in the plane of $\eta,n$. Also shown by crosses are the values of $\eta,n$ that mark the boundary between the region of stability (for all values of $\dP$) at high $n$ from that of instability (for some values of $\dP$) at low $n$.

As is evident from Figs.~\ref{fig:WM_map1} to ~\ref{fig:WM_map2}, the results are largely independent of the solvent viscosity $\eta$ in the White-Metzner model. This should not be surprising: the solvent contribution to the flow curve stress, $\Sigma=\gdot|\gdot|^{n-1}+\eta\gdot$, is much smaller relative to the viscoelastic one, for realistic values of $\eta<G\tau=1$, at least in the window of stresses explored across the channel in the vicinity of instability onset. In this way, the collapse obtained with $\eta$ in Figs.~\ref{fig:WM_map1} and~\ref{fig:WM_criterion} is relatively trivial. The more interesting feature of these results is the fact that the instability threshold appears to follow the same functional dependency on $n$, albeit with different values of the fitting parameters $\alpha_0,\alpha_1,\alpha_2$.

\section{Conclusions}
\label{sec:conclusions}

We have combined linear stability analysis with full numerical simulation to study pressure driven channel flow within three widely used constitutive models of shear thinning viscoelastic flow: the microscopically motivated Rolie-Poly model, and the phenomenological Johnson-Segalman and White-Metzner models. We have shown an initially 1D base state to be linearly unstable to the onset of 2D perturbations with wavevector in the flow direction, in all three models, for a high enough degree of shear thinning. That we indeed find instability across several constitutive models suggests that the instability may be generic across shear thinning polymeric fluids. Within each model, we have calculated the minimal degree of shear thinning  needed to observe instability, finding instability below a critical value $n^*$ of the logarithmic slope of the flow curve at its shallowest point, $n=d\log\Sigma/d\log\gdot|_{\rm min}$, with $n^*\approx 0.2$ in Rolie-Poly, $n^*\approx 0.11$ in Johnson-Segalman, and $n^*\approx 0.3$ in White-Metzner.

Within each of the Rolie-Poly, Johnson-Segalman and White-Metzner models, we have shown the critical adimensional pressure drop at the onset of instability  to follow, to a reasonable level of approximation, a scaling function expressed in terms of (i) this degree of shear thinning, $n$, and (ii) the maximum derivative of the the first normal stress with respect to the shear stress. In each model, the scaling function is of the same functional form. However, it is characterised by three fitting parameters, $\alpha_0,\alpha_1,\alpha_2$, the values of which differ across the three models. In this way, the onset criterion obtained here is clearly only a partial success. An open challenge for future work is to recast these fitting parameters in terms of any additional relevant dimensionless constitutive properties, leading to a re-expressed set of fitting parameters that have the same values across all models, thereby providing a truly universal criterion for the onset of instability. This goal might be unattainable, however.

In the Rolie-Poly and Johnson-Segalman models, we have given evidence suggesting that the mechanism of instability involves the presence of a quasi-interface at some location in each half of the channel, across which the first normal stress and shear rate both vary steeply. In this way, the instability appears similar to that arising at the interface between layered fluids or shear bands. In the White-Metzner model, no such quasi-interface exists. An unresolved puzzle is why the same criterion for instability (albeit with different values of the fitting parameters) appears to hold in the  White-Metzner model as in the Rolie-Poly and Johnson-Segalman models.


Recently, the original predictions of instability within the White-Metzner model~\cite{wilson1999instabilityWM} were extended by including an additional stress contribution that matches the shear thinning of White-Metzner, but that responds instantaneously to the imposed flow, and lacks any normal stress contribution~\cite{Castillo2017}. As the contribution from this term increased relative to the viscoelastic contribution from White-Metzner, the level of instability decreased and stability was eventually restored.
The overall scaling of the modulus and relaxation time of the White-Metzner model were also respectively decreased and increased with the increasing additional stress contribution. The decrease of the modulus, in particular, decreases the normal stress. This work appears consistent with our finding, of increasing levels of stability with decreasing levels of normal stress. In future work, it would be interesting to see whether the predictions of~\cite{Castillo2017} indeed accord with our criterion.

As things currently stand, then, two possible mechanisms have been identified for instability: a jump in $N_1$ across an interface between two fluid layers~\cite{wilson1997short}, and an instability of shear thinning flow, even if $N_1$ varies smoothly across the channel~\cite{wilson1999instabilityWM}. This raises the question of whether, in the various models considered here, the instabilities are related primarily to rapid variations in $N_1$ or in $\gdot$ across the channel. The answer to this remains unclear at present, but it is worth collecting the evidence for the spatial nature of the eigenfunctions in the different models. In the Rolie-Poly and Johnson-Segalman models, both the viscosity and $N_1$ vary rapidly in the quasi-interface region, and the eigenfunction is localised here. In the White-Metzner model~\cite{wilson1999instabilityWM}, the viscosity and $N_1$ both vary as power laws across the channel, and the eigenfunction appears delocalised right across the channel. In a very recent study~\cite{Castillo2018} of a model in which $N_1$ varies smoothly across the channel but the viscosity varies relatively rapidly near the walls, the eigenfunction was found to relatively localised near the walls. 
Refs~\cite{Miller2007,Miller2007a} considered layered viscoelastic fluids with matched viscosities, but different relaxation times and so normal stresses. The interface was found to be unstable, even with the addition of a surface tension. As things stand, therefore, it remains unclear whether instabilities of the kind studied in this work are generically driven primarily by variations across the channel in the base state viscosity, in the base state normal stress, or even in some dynamical property not captured by the stationary underlying base state.

We now make some comments about the Giesekus model, for which instability was previously found in Ref.~\cite{grillet2002stability}, and for which we have performed here a more modest range of numerical explorations. The stability behaviour of this model is rather different from that of the other three models. In particular, the maximum logarithmic slope $n^*$ of the flow curve (minimised across the flow curve) that permits instability is much smaller in this model, and furthermore depends strongly on the solvent viscosity. For values of $\eta$ that we explored in the range $0$ to $0.015$, $n^*$ appears to vary in the range $0$ - $0.04$. Therefore, a fluid must be much more highly shear thinning to be predicted unstable by Giesekus than by the other three models. Furthermore, for any set of values of $\alpha,\eta$, the window of pressure drops for which instability arises (if at all) is very much narrower than in the other models, with re-entrant stability following at a pressure drop not much greater than that for instability onset. Taken overall, then, the Giesekus model appears to predict a much greater level stability in channel flow. It is probably just a coincidence, although worth remarking, that the same model also predicted much greater levels of stability against the formation of transient bands in shear startup (compared with the Rolie-Poly model) in Ref.~\cite{moorcroft2014shear}. 

An aspect of the Giesekus model noted in Ref.\cite{giesekus1982simple} is that for a non-linear relaxation parameter $\alpha=1$, the shear and normal stress
constitutive curves exactly match those of the co-rotational Maxwell model, which is equivalent to Johnson Segalman for $a=0$. In this way, the 1D base state of the Giesekus model at $\alpha=1$ is the same as that of the Johnson-Segalman model at $a=0$. 
Puzzlingly, the former appears stable, within Giesekus dynamics (data not shown), while the latter appears unstable, within the Johnson-Segalman model (recall Fig.~\ref{fig:JS_onset}). This observation again suggests that some feature of constitutive dynamics not represented by the stationary base state plays a role in determining instability.

All the constitutive models used in this study are designed to describe shear thinning viscoelastic fluids. Given that they predict rather different stability properties from each other (quantitatively for the Rolie-Poly, Johnson-Segalman and White-Metzner model and even qualitatively for the Giesekus model), this suggests that prediction of stability properties could be used as a helpful constraint in model building, beyond the fitting to homogeneous rheological functions as is more usually carried out.

Finally, it is worth remarking that we have considered here a time-independent and x-independent 1D base state, about which 2D perturbations may or may not grow. In practice, one may instead be interested in whether a 1D state of fluid flowing into and out of a channel of a finite length becomes unstable: in particular, in whether the instability has time to develop for a channel of any given length. To answer that question fully, one would have to simulate that more dynamic situation. A reasonable estimate may, however, be gained from our results by comparing the timescale $\omega^{-1}$ for instability to develop with the residence time in the channel, $L_x/V$, where $V$ is the velocity at the channel midpoint.

{\em Acknowledgements ---} We thank Mike Graham, Hugo Castillo-Sanchez and Helen Wilson for useful discussions. We also acknowledge the SOFI CDT, Durham University, EPSRC (EP/L015536/1), the European Research
Council under the European Union Seventh Framework
Programme (FP7/2007-2013)/	ERC	grant	agreement	number	279365, and Schlumberger Cambridge Research for funding.

\bibliography{ChannelFlow}

\begin{thebibliography}{38}%
\makeatletter
\providecommand \@ifxundefined [1]{%
 \@ifx{#1\undefined}
}%
\providecommand \@ifnum [1]{%
 \ifnum #1\expandafter \@firstoftwo
 \else \expandafter \@secondoftwo
 \fi
}%
\providecommand \@ifx [1]{%
 \ifx #1\expandafter \@firstoftwo
 \else \expandafter \@secondoftwo
 \fi
}%
\providecommand \natexlab [1]{#1}%
\providecommand \enquote  [1]{``#1''}%
\providecommand \bibnamefont  [1]{#1}%
\providecommand \bibfnamefont [1]{#1}%
\providecommand \citenamefont [1]{#1}%
\providecommand \href@noop [0]{\@secondoftwo}%
\providecommand \href [0]{\begingroup \@sanitize@url \@href}%
\providecommand \@href[1]{\@@startlink{#1}\@@href}%
\providecommand \@@href[1]{\endgroup#1\@@endlink}%
\providecommand \@sanitize@url [0]{\catcode `\\12\catcode `\$12\catcode
  `\&12\catcode `\#12\catcode `\^12\catcode `\_12\catcode `\%12\relax}%
\providecommand \@@startlink[1]{}%
\providecommand \@@endlink[0]{}%
\providecommand \url  [0]{\begingroup\@sanitize@url \@url }%
\providecommand \@url [1]{\endgroup\@href {#1}{\urlprefix }}%
\providecommand \urlprefix  [0]{URL }%
\providecommand \Eprint [0]{\href }%
\providecommand \doibase [0]{http://dx.doi.org/}%
\providecommand \selectlanguage [0]{\@gobble}%
\providecommand \bibinfo  [0]{\@secondoftwo}%
\providecommand \bibfield  [0]{\@secondoftwo}%
\providecommand \translation [1]{[#1]}%
\providecommand \BibitemOpen [0]{}%
\providecommand \bibitemStop [0]{}%
\providecommand \bibitemNoStop [0]{.\EOS\space}%
\providecommand \EOS [0]{\spacefactor3000\relax}%
\providecommand \BibitemShut  [1]{\csname bibitem#1\endcsname}%
\let\auto@bib@innerbib\@empty
\bibitem [{\citenamefont {Larson}(1992)}]{Larson1992}%
  \BibitemOpen
  \bibfield  {author} {\bibinfo {author} {\bibfnamefont {R.~G.}\ \bibnamefont
  {Larson}},\ }\href {\doibase 10.1007/BF00366504} {\bibfield  {journal}
  {\bibinfo  {journal} {Rheologica Acta}\ }\textbf {\bibinfo {volume} {31}},\
  \bibinfo {pages} {213} (\bibinfo {year} {1992})}\BibitemShut {NoStop}%
\bibitem [{\citenamefont {Larson}\ \emph {et~al.}(1990)\citenamefont {Larson},
  \citenamefont {Shaqfeh},\ and\ \citenamefont {Muller}}]{larson1990purely}%
  \BibitemOpen
  \bibfield  {author} {\bibinfo {author} {\bibfnamefont {R.~G.}\ \bibnamefont
  {Larson}}, \bibinfo {author} {\bibfnamefont {E.~S.}\ \bibnamefont {Shaqfeh}},
  \ and\ \bibinfo {author} {\bibfnamefont {S.~J.}\ \bibnamefont {Muller}},\
  }\href@noop {} {\bibfield  {journal} {\bibinfo  {journal} {Journal of Fluid
  Mechanics}\ }\textbf {\bibinfo {volume} {218}},\ \bibinfo {pages} {573}
  (\bibinfo {year} {1990})}\BibitemShut {NoStop}%
\bibitem [{\citenamefont {Olagunju}(1995)}]{olagunju1995instabilities}%
  \BibitemOpen
  \bibfield  {author} {\bibinfo {author} {\bibfnamefont {D.~O.}\ \bibnamefont
  {Olagunju}},\ }\href@noop {} {\bibfield  {journal} {\bibinfo  {journal}
  {Zeitschrift f{\"u}r angewandte Mathematik und Physik ZAMP}\ }\textbf
  {\bibinfo {volume} {46}},\ \bibinfo {pages} {224} (\bibinfo {year}
  {1995})}\BibitemShut {NoStop}%
\bibitem [{\citenamefont {Schiamberg}\ \emph {et~al.}(2006)\citenamefont
  {Schiamberg}, \citenamefont {Shereda}, \citenamefont {Hu},\ and\
  \citenamefont {Larson}}]{schiamberg2006transitional}%
  \BibitemOpen
  \bibfield  {author} {\bibinfo {author} {\bibfnamefont {B.~A.}\ \bibnamefont
  {Schiamberg}}, \bibinfo {author} {\bibfnamefont {L.~T.}\ \bibnamefont
  {Shereda}}, \bibinfo {author} {\bibfnamefont {H.}~\bibnamefont {Hu}}, \ and\
  \bibinfo {author} {\bibfnamefont {R.~G.}\ \bibnamefont {Larson}},\
  }\href@noop {} {\bibfield  {journal} {\bibinfo  {journal} {Journal of Fluid
  Mechanics}\ }\textbf {\bibinfo {volume} {554}},\ \bibinfo {pages} {191}
  (\bibinfo {year} {2006})}\BibitemShut {NoStop}%
\bibitem [{\citenamefont {Pakdel}\ and\ \citenamefont
  {McKinley}(1996)}]{pakdel1996elastic}%
  \BibitemOpen
  \bibfield  {author} {\bibinfo {author} {\bibfnamefont {P.}~\bibnamefont
  {Pakdel}}\ and\ \bibinfo {author} {\bibfnamefont {G.~H.}\ \bibnamefont
  {McKinley}},\ }\href@noop {} {\bibfield  {journal} {\bibinfo  {journal}
  {Physical Review Letters}\ }\textbf {\bibinfo {volume} {77}},\ \bibinfo
  {pages} {2459} (\bibinfo {year} {1996})}\BibitemShut {NoStop}%
\bibitem [{\citenamefont {Wilson}\ \emph {et~al.}(1999)\citenamefont {Wilson},
  \citenamefont {Renardy},\ and\ \citenamefont
  {Renardy}}]{wilson1999structure}%
  \BibitemOpen
  \bibfield  {author} {\bibinfo {author} {\bibfnamefont {H.~J.}\ \bibnamefont
  {Wilson}}, \bibinfo {author} {\bibfnamefont {M.}~\bibnamefont {Renardy}}, \
  and\ \bibinfo {author} {\bibfnamefont {Y.}~\bibnamefont {Renardy}},\
  }\href@noop {} {\bibfield  {journal} {\bibinfo  {journal} {Journal of
  non-newtonian fluid mechanics}\ }\textbf {\bibinfo {volume} {80}},\ \bibinfo
  {pages} {251} (\bibinfo {year} {1999})}\BibitemShut {NoStop}%
\bibitem [{\citenamefont {Ho}\ and\ \citenamefont
  {Denn}(1977)}]{ho1977stability}%
  \BibitemOpen
  \bibfield  {author} {\bibinfo {author} {\bibfnamefont {T.~C.}\ \bibnamefont
  {Ho}}\ and\ \bibinfo {author} {\bibfnamefont {M.~M.}\ \bibnamefont {Denn}},\
  }\href@noop {} {\bibfield  {journal} {\bibinfo  {journal} {Journal of
  Non-Newtonian Fluid Mechanics}\ }\textbf {\bibinfo {volume} {3}},\ \bibinfo
  {pages} {179} (\bibinfo {year} {1977})}\BibitemShut {NoStop}%
\bibitem [{\citenamefont {Morozov}\ and\ \citenamefont {van
  Saarloos}(2005)}]{morozov2005subcritical}%
  \BibitemOpen
  \bibfield  {author} {\bibinfo {author} {\bibfnamefont {A.~N.}\ \bibnamefont
  {Morozov}}\ and\ \bibinfo {author} {\bibfnamefont {W.}~\bibnamefont {van
  Saarloos}},\ }\href@noop {} {\bibfield  {journal} {\bibinfo  {journal}
  {Physical review letters}\ }\textbf {\bibinfo {volume} {95}},\ \bibinfo
  {pages} {024501} (\bibinfo {year} {2005})}\BibitemShut {NoStop}%
\bibitem [{\citenamefont {Qin}\ and\ \citenamefont
  {Arratia}(2017)}]{qin2017characterizing}%
  \BibitemOpen
  \bibfield  {author} {\bibinfo {author} {\bibfnamefont {B.}~\bibnamefont
  {Qin}}\ and\ \bibinfo {author} {\bibfnamefont {P.~E.}\ \bibnamefont
  {Arratia}},\ }\href@noop {} {\bibfield  {journal} {\bibinfo  {journal}
  {Physical Review Fluids}\ }\textbf {\bibinfo {volume} {2}},\ \bibinfo {pages}
  {083302} (\bibinfo {year} {2017})}\BibitemShut {NoStop}%
\bibitem [{\citenamefont {Pan}\ \emph {et~al.}(2013)\citenamefont {Pan},
  \citenamefont {Morozov}, \citenamefont {Wagner},\ and\ \citenamefont
  {Arratia}}]{pan2013nonlinear}%
  \BibitemOpen
  \bibfield  {author} {\bibinfo {author} {\bibfnamefont {L.}~\bibnamefont
  {Pan}}, \bibinfo {author} {\bibfnamefont {A.}~\bibnamefont {Morozov}},
  \bibinfo {author} {\bibfnamefont {C.}~\bibnamefont {Wagner}}, \ and\ \bibinfo
  {author} {\bibfnamefont {P.}~\bibnamefont {Arratia}},\ }\href@noop {}
  {\bibfield  {journal} {\bibinfo  {journal} {Physical review letters}\
  }\textbf {\bibinfo {volume} {110}},\ \bibinfo {pages} {174502} (\bibinfo
  {year} {2013})}\BibitemShut {NoStop}%
\bibitem [{\citenamefont {Wilson}\ and\ \citenamefont
  {Rallison}(1999)}]{wilson1999instabilityWM}%
  \BibitemOpen
  \bibfield  {author} {\bibinfo {author} {\bibfnamefont {H.~J.}\ \bibnamefont
  {Wilson}}\ and\ \bibinfo {author} {\bibfnamefont {J.~M.}\ \bibnamefont
  {Rallison}},\ }\href@noop {} {\bibfield  {journal} {\bibinfo  {journal}
  {Journal of non-newtonian fluid mechanics}\ }\textbf {\bibinfo {volume}
  {87}},\ \bibinfo {pages} {75} (\bibinfo {year} {1999})}\BibitemShut {NoStop}%
\bibitem [{\citenamefont {Wilson}\ and\ \citenamefont
  {Loridan}(2015)}]{Wilson2015}%
  \BibitemOpen
  \bibfield  {author} {\bibinfo {author} {\bibfnamefont {H.~J.}\ \bibnamefont
  {Wilson}}\ and\ \bibinfo {author} {\bibfnamefont {V.}~\bibnamefont
  {Loridan}},\ }\href {\doibase 10.1016/j.jnnfm.2015.07.002} {\bibfield
  {journal} {\bibinfo  {journal} {Journal of Non-Newtonian Fluid Mechanics}\ }
  (\bibinfo {year} {2015}),\ 10.1016/j.jnnfm.2015.07.002}\BibitemShut {NoStop}%
\bibitem [{\citenamefont {Castillo}\ and\ \citenamefont
  {Wilson}(2017)}]{Castillo2017}%
  \BibitemOpen
  \bibfield  {author} {\bibinfo {author} {\bibfnamefont {H.~A.}\ \bibnamefont
  {Castillo}}\ and\ \bibinfo {author} {\bibfnamefont {H.~J.}\ \bibnamefont
  {Wilson}},\ }\href {\doibase 10.1016/j.jnnfm.2017.06.001} {\bibfield
  {journal} {\bibinfo  {journal} {Journal of Non-Newtonian Fluid Mechanics}\ }
  (\bibinfo {year} {2017}),\ 10.1016/j.jnnfm.2017.06.001}\BibitemShut {NoStop}%
\bibitem [{\citenamefont {Grillet}\ \emph {et~al.}(2002)\citenamefont
  {Grillet}, \citenamefont {Bogaerds}, \citenamefont {Peters},\ and\
  \citenamefont {Baaijens}}]{grillet2002stability}%
  \BibitemOpen
  \bibfield  {author} {\bibinfo {author} {\bibfnamefont {A.~M.}\ \bibnamefont
  {Grillet}}, \bibinfo {author} {\bibfnamefont {A.~C.}\ \bibnamefont
  {Bogaerds}}, \bibinfo {author} {\bibfnamefont {G.~W.}\ \bibnamefont
  {Peters}}, \ and\ \bibinfo {author} {\bibfnamefont {F.~P.}\ \bibnamefont
  {Baaijens}},\ }\href@noop {} {\bibfield  {journal} {\bibinfo  {journal}
  {Journal of non-newtonian fluid mechanics}\ }\textbf {\bibinfo {volume}
  {103}},\ \bibinfo {pages} {221} (\bibinfo {year} {2002})}\BibitemShut
  {NoStop}%
\bibitem [{\citenamefont {Castillo}\ and\ \citenamefont
  {Wilson}(2018)}]{Castillo2018}%
  \BibitemOpen
  \bibfield  {author} {\bibinfo {author} {\bibfnamefont {H.~A.}\ \bibnamefont
  {Castillo}}\ and\ \bibinfo {author} {\bibfnamefont {H.~J.}\ \bibnamefont
  {Wilson}},\ }\href {\doibase https://doi.org/10.1016/j.jnnfm.2018.07.009}
  {\bibfield  {journal} {\bibinfo  {journal} {Journal of Non-Newtonian Fluid
  Mechanics}\ }\textbf {\bibinfo {volume} {261}},\ \bibinfo {pages} {10 }
  (\bibinfo {year} {2018})}\BibitemShut {NoStop}%
\bibitem [{\citenamefont {Picaut}\ \emph {et~al.}(2017)\citenamefont {Picaut},
  \citenamefont {Ronsin}, \citenamefont {Caroli},\ and\ \citenamefont
  {Baumberger}}]{Picaut2017}%
  \BibitemOpen
  \bibfield  {author} {\bibinfo {author} {\bibfnamefont {L.}~\bibnamefont
  {Picaut}}, \bibinfo {author} {\bibfnamefont {O.}~\bibnamefont {Ronsin}},
  \bibinfo {author} {\bibfnamefont {C.}~\bibnamefont {Caroli}}, \ and\ \bibinfo
  {author} {\bibfnamefont {T.}~\bibnamefont {Baumberger}},\ }\href {\doibase
  10.1103/PhysRevFluids.2.083303} {\bibfield  {journal} {\bibinfo  {journal}
  {Physical Review Fluids}\ } (\bibinfo {year} {2017}),\
  10.1103/PhysRevFluids.2.083303}\BibitemShut {NoStop}%
\bibitem [{\citenamefont {Bodiguel}\ \emph {et~al.}(2015)\citenamefont
  {Bodiguel}, \citenamefont {Beaumont}, \citenamefont {Machado}, \citenamefont
  {Martinie}, \citenamefont {Kellay},\ and\ \citenamefont
  {Colin}}]{Bodiguel2015}%
  \BibitemOpen
  \bibfield  {author} {\bibinfo {author} {\bibfnamefont {H.}~\bibnamefont
  {Bodiguel}}, \bibinfo {author} {\bibfnamefont {J.}~\bibnamefont {Beaumont}},
  \bibinfo {author} {\bibfnamefont {A.}~\bibnamefont {Machado}}, \bibinfo
  {author} {\bibfnamefont {L.}~\bibnamefont {Martinie}}, \bibinfo {author}
  {\bibfnamefont {H.}~\bibnamefont {Kellay}}, \ and\ \bibinfo {author}
  {\bibfnamefont {A.}~\bibnamefont {Colin}},\ }\href {\doibase
  10.1103/PhysRevLett.114.028302} {\bibfield  {journal} {\bibinfo  {journal}
  {Physical Review Letters}\ } (\bibinfo {year} {2015}),\
  10.1103/PhysRevLett.114.028302}\BibitemShut {NoStop}%
\bibitem [{\citenamefont {Poole}(2016)}]{Poole2016}%
  \BibitemOpen
  \bibfield  {author} {\bibinfo {author} {\bibfnamefont {R.~J.}\ \bibnamefont
  {Poole}},\ }\href {\doibase 10.1103/PhysRevFluids.1.041301} {\bibfield
  {journal} {\bibinfo  {journal} {Physical Review Fluids}\ }\textbf {\bibinfo
  {volume} {1}},\ \bibinfo {pages} {041301} (\bibinfo {year}
  {2016})}\BibitemShut {NoStop}%
\bibitem [{\citenamefont {Bonn}\ \emph {et~al.}(2011)\citenamefont {Bonn},
  \citenamefont {Ingremeau}, \citenamefont {Amarouchene},\ and\ \citenamefont
  {Kellay}}]{bonn2011large}%
  \BibitemOpen
  \bibfield  {author} {\bibinfo {author} {\bibfnamefont {D.}~\bibnamefont
  {Bonn}}, \bibinfo {author} {\bibfnamefont {F.}~\bibnamefont {Ingremeau}},
  \bibinfo {author} {\bibfnamefont {Y.}~\bibnamefont {Amarouchene}}, \ and\
  \bibinfo {author} {\bibfnamefont {H.}~\bibnamefont {Kellay}},\ }\href@noop {}
  {\bibfield  {journal} {\bibinfo  {journal} {Physical Review E}\ }\textbf
  {\bibinfo {volume} {84}},\ \bibinfo {pages} {045301} (\bibinfo {year}
  {2011})}\BibitemShut {NoStop}%
\bibitem [{\citenamefont {Denn}(2001)}]{denn2001extrusion}%
  \BibitemOpen
  \bibfield  {author} {\bibinfo {author} {\bibfnamefont {M.~M.}\ \bibnamefont
  {Denn}},\ }\href@noop {} {\bibfield  {journal} {\bibinfo  {journal} {Annual
  Review of Fluid Mechanics}\ }\textbf {\bibinfo {volume} {33}},\ \bibinfo
  {pages} {265} (\bibinfo {year} {2001})}\BibitemShut {NoStop}%
\bibitem [{\citenamefont {Likhtman}\ and\ \citenamefont
  {Graham}(2003)}]{Likhtman2003}%
  \BibitemOpen
  \bibfield  {author} {\bibinfo {author} {\bibfnamefont {A.~E.}\ \bibnamefont
  {Likhtman}}\ and\ \bibinfo {author} {\bibfnamefont {R.~S.}\ \bibnamefont
  {Graham}},\ }\href {\doibase 10.1016/S0377-0257(03)00114-9} {\bibfield
  {journal} {\bibinfo  {journal} {Journal of Non-Newtonian Fluid Mechanics}\ }
  (\bibinfo {year} {2003}),\ 10.1016/S0377-0257(03)00114-9}\BibitemShut
  {NoStop}%
\bibitem [{\citenamefont {Johnson}\ and\ \citenamefont
  {Segalman}(1977)}]{Johnson1977}%
  \BibitemOpen
  \bibfield  {author} {\bibinfo {author} {\bibfnamefont {M.~W.}\ \bibnamefont
  {Johnson}}\ and\ \bibinfo {author} {\bibfnamefont {D.}~\bibnamefont
  {Segalman}},\ }\href {\doibase 10.1016/0377-0257(77)80003-7} {\enquote
  {\bibinfo {title} {{A model for viscoelastic fluid behavior which allows
  non-affine deformation}},}\ } (\bibinfo {year} {1977})\BibitemShut {NoStop}%
\bibitem [{\citenamefont {White}\ and\ \citenamefont
  {Metzner}(1963)}]{white1963development}%
  \BibitemOpen
  \bibfield  {author} {\bibinfo {author} {\bibfnamefont {J.}~\bibnamefont
  {White}}\ and\ \bibinfo {author} {\bibfnamefont {A.}~\bibnamefont
  {Metzner}},\ }\href@noop {} {\bibfield  {journal} {\bibinfo  {journal}
  {Journal of Applied Polymer Science}\ }\textbf {\bibinfo {volume} {7}},\
  \bibinfo {pages} {1867} (\bibinfo {year} {1963})}\BibitemShut {NoStop}%
\bibitem [{\citenamefont {Giesekus}(1982)}]{giesekus1982simple}%
  \BibitemOpen
  \bibfield  {author} {\bibinfo {author} {\bibfnamefont {H.}~\bibnamefont
  {Giesekus}},\ }\href@noop {} {\bibfield  {journal} {\bibinfo  {journal}
  {Journal of Non-Newtonian Fluid Mechanics}\ }\textbf {\bibinfo {volume}
  {11}},\ \bibinfo {pages} {69} (\bibinfo {year} {1982})}\BibitemShut {NoStop}%
\bibitem [{\citenamefont {Wilson}\ and\ \citenamefont
  {Rallison}(1997)}]{wilson1997short}%
  \BibitemOpen
  \bibfield  {author} {\bibinfo {author} {\bibfnamefont {H.~J.}\ \bibnamefont
  {Wilson}}\ and\ \bibinfo {author} {\bibfnamefont {J.~M.}\ \bibnamefont
  {Rallison}},\ }\href@noop {} {\bibfield  {journal} {\bibinfo  {journal}
  {Journal of non-newtonian fluid mechanics}\ }\textbf {\bibinfo {volume}
  {72}},\ \bibinfo {pages} {237} (\bibinfo {year} {1997})}\BibitemShut
  {NoStop}%
\bibitem [{\citenamefont {Fielding}(2005)}]{fielding2005linear}%
  \BibitemOpen
  \bibfield  {author} {\bibinfo {author} {\bibfnamefont {S.~M.}\ \bibnamefont
  {Fielding}},\ }\href@noop {} {\bibfield  {journal} {\bibinfo  {journal}
  {Physical review letters}\ }\textbf {\bibinfo {volume} {95}},\ \bibinfo
  {pages} {134501} (\bibinfo {year} {2005})}\BibitemShut {NoStop}%
\bibitem [{\citenamefont {Fielding}\ and\ \citenamefont
  {Olmsted}(2006)}]{fielding2006nonlinear}%
  \BibitemOpen
  \bibfield  {author} {\bibinfo {author} {\bibfnamefont {S.~M.}\ \bibnamefont
  {Fielding}}\ and\ \bibinfo {author} {\bibfnamefont {P.~D.}\ \bibnamefont
  {Olmsted}},\ }\href@noop {} {\bibfield  {journal} {\bibinfo  {journal}
  {Physical review letters}\ }\textbf {\bibinfo {volume} {96}},\ \bibinfo
  {pages} {104502} (\bibinfo {year} {2006})}\BibitemShut {NoStop}%
\bibitem [{\citenamefont {Fielding}\ and\ \citenamefont
  {Wilson}(2010)}]{fielding2010shear}%
  \BibitemOpen
  \bibfield  {author} {\bibinfo {author} {\bibfnamefont {S.~M.}\ \bibnamefont
  {Fielding}}\ and\ \bibinfo {author} {\bibfnamefont {H.~J.}\ \bibnamefont
  {Wilson}},\ }\href@noop {} {\bibfield  {journal} {\bibinfo  {journal}
  {Journal of Non-Newtonian Fluid Mechanics}\ }\textbf {\bibinfo {volume}
  {165}},\ \bibinfo {pages} {196} (\bibinfo {year} {2010})}\BibitemShut
  {NoStop}%
\bibitem [{\citenamefont {Larson}(2013)}]{larson2013constitutive}%
  \BibitemOpen
  \bibfield  {author} {\bibinfo {author} {\bibfnamefont {R.~G.}\ \bibnamefont
  {Larson}},\ }\href@noop {} {\emph {\bibinfo {title} {Constitutive Equations
  for Polymer Melts and Solutions: Butterworths Series in Chemical
  Engineering}}}\ (\bibinfo  {publisher} {Butterworth-Heinemann},\ \bibinfo
  {year} {2013})\BibitemShut {NoStop}%
\bibitem [{\citenamefont {Doi}\ and\ \citenamefont
  {Edwards}(1988)}]{doi1988theory}%
  \BibitemOpen
  \bibfield  {author} {\bibinfo {author} {\bibfnamefont {M.}~\bibnamefont
  {Doi}}\ and\ \bibinfo {author} {\bibfnamefont {S.~F.}\ \bibnamefont
  {Edwards}},\ }\href@noop {} {\emph {\bibinfo {title} {The theory of polymer
  dynamics}}},\ Vol.~\bibinfo {volume} {73}\ (\bibinfo  {publisher} {oxford
  university press},\ \bibinfo {year} {1988})\BibitemShut {NoStop}%
\bibitem [{\citenamefont {Graham}\ \emph {et~al.}(2003)\citenamefont {Graham},
  \citenamefont {Likhtman}, \citenamefont {McLeish},\ and\ \citenamefont
  {Milner}}]{Graham2003}%
  \BibitemOpen
  \bibfield  {author} {\bibinfo {author} {\bibfnamefont {R.~S.}\ \bibnamefont
  {Graham}}, \bibinfo {author} {\bibfnamefont {A.~E.}\ \bibnamefont
  {Likhtman}}, \bibinfo {author} {\bibfnamefont {T.~C.~B.}\ \bibnamefont
  {McLeish}}, \ and\ \bibinfo {author} {\bibfnamefont {S.~T.}\ \bibnamefont
  {Milner}},\ }\href {\doibase 10.1122/1.1595099} {\bibfield  {journal}
  {\bibinfo  {journal} {Journal of Rheology}\ } (\bibinfo {year} {2003}),\
  10.1122/1.1595099}\BibitemShut {NoStop}%
\bibitem [{\citenamefont {Ianniruberto}\ and\ \citenamefont
  {Marrucci}(2014)}]{ianniruberto2014convective}%
  \BibitemOpen
  \bibfield  {author} {\bibinfo {author} {\bibfnamefont {G.}~\bibnamefont
  {Ianniruberto}}\ and\ \bibinfo {author} {\bibfnamefont {G.}~\bibnamefont
  {Marrucci}},\ }\href@noop {} {\bibfield  {journal} {\bibinfo  {journal}
  {Journal of Rheology}\ }\textbf {\bibinfo {volume} {58}},\ \bibinfo {pages}
  {89} (\bibinfo {year} {2014})}\BibitemShut {NoStop}%
\bibitem [{\citenamefont {Spiess}(1987)}]{spiess1987rb}%
  \BibitemOpen
  \bibfield  {author} {\bibinfo {author} {\bibfnamefont {H.}~\bibnamefont
  {Spiess}},\ }\href@noop {} {\bibfield  {journal} {\bibinfo  {journal}
  {Berichte der Bunsengesellschaft f{\"u}r physikalische Chemie}\ }\textbf
  {\bibinfo {volume} {91}},\ \bibinfo {pages} {1397} (\bibinfo {year}
  {1987})}\BibitemShut {NoStop}%
\bibitem [{\citenamefont {Lockett}(1969)}]{LOCKETT1969337}%
  \BibitemOpen
  \bibfield  {author} {\bibinfo {author} {\bibfnamefont {F.}~\bibnamefont
  {Lockett}},\ }\href {\doibase https://doi.org/10.1016/0020-7225(69)90044-5}
  {\bibfield  {journal} {\bibinfo  {journal} {International Journal of
  Engineering Science}\ }\textbf {\bibinfo {volume} {7}},\ \bibinfo {pages}
  {337 } (\bibinfo {year} {1969})}\BibitemShut {NoStop}%
\bibitem [{\citenamefont {Lu}\ \emph {et~al.}(2000)\citenamefont {Lu},
  \citenamefont {Olmsted},\ and\ \citenamefont {Ball}}]{lu2000effects}%
  \BibitemOpen
  \bibfield  {author} {\bibinfo {author} {\bibfnamefont {C.-Y.~D.}\
  \bibnamefont {Lu}}, \bibinfo {author} {\bibfnamefont {P.~D.}\ \bibnamefont
  {Olmsted}}, \ and\ \bibinfo {author} {\bibfnamefont {R.}~\bibnamefont
  {Ball}},\ }\href@noop {} {\bibfield  {journal} {\bibinfo  {journal} {Physical
  Review Letters}\ }\textbf {\bibinfo {volume} {84}},\ \bibinfo {pages} {642}
  (\bibinfo {year} {2000})}\BibitemShut {NoStop}%
\bibitem [{\citenamefont {Miller}\ and\ \citenamefont
  {Rallison}(2007{\natexlab{a}})}]{Miller2007}%
  \BibitemOpen
  \bibfield  {author} {\bibinfo {author} {\bibfnamefont {J.~C.}\ \bibnamefont
  {Miller}}\ and\ \bibinfo {author} {\bibfnamefont {J.~M.}\ \bibnamefont
  {Rallison}},\ }\href {\doibase 10.1016/j.jnnfm.2007.01.008} {\bibfield
  {journal} {\bibinfo  {journal} {Journal of Non-Newtonian Fluid Mechanics}\ }
  (\bibinfo {year} {2007}{\natexlab{a}}),\
  10.1016/j.jnnfm.2007.01.008}\BibitemShut {NoStop}%
\bibitem [{\citenamefont {Miller}\ and\ \citenamefont
  {Rallison}(2007{\natexlab{b}})}]{Miller2007a}%
  \BibitemOpen
  \bibfield  {author} {\bibinfo {author} {\bibfnamefont {J.~C.}\ \bibnamefont
  {Miller}}\ and\ \bibinfo {author} {\bibfnamefont {J.~M.}\ \bibnamefont
  {Rallison}},\ }\href {\doibase 10.1016/j.jnnfm.2007.01.009} {\bibfield
  {journal} {\bibinfo  {journal} {Journal of Non-Newtonian Fluid Mechanics}\ }
  (\bibinfo {year} {2007}{\natexlab{b}}),\
  10.1016/j.jnnfm.2007.01.009}\BibitemShut {NoStop}%
\bibitem [{\citenamefont {Moorcroft}\ and\ \citenamefont
  {Fielding}(2014)}]{moorcroft2014shear}%
  \BibitemOpen
  \bibfield  {author} {\bibinfo {author} {\bibfnamefont {R.~L.}\ \bibnamefont
  {Moorcroft}}\ and\ \bibinfo {author} {\bibfnamefont {S.~M.}\ \bibnamefont
  {Fielding}},\ }\href@noop {} {\bibfield  {journal} {\bibinfo  {journal}
  {Journal of Rheology}\ }\textbf {\bibinfo {volume} {58}},\ \bibinfo {pages}
  {103} (\bibinfo {year} {2014})}\BibitemShut {NoStop}%
\end{thebibliography}%
\bibliographystyle{apsrev4-1}

\newpage

\newpage

\end{document}